# Ionic switching between supermagnetic states

Sumaiyatul Ahsan[1], Yin-Ju Yen[2], Andrew H. Comstock[3], Tanya A. Balandin[1], Jason D. Azoulay[1,3], Jamie H. Warner[2], Dario A. Arena[4], Faisal M. Alamgir[1]*

**Affiliations:**

[1]Department of Materials Science and Engineering; Georgia Institute of Technology, Atlanta, GA 30332, USA
[2]Department of Mechanical Engineering; University of Texas at Austin, TX 78712, USA
[3]Department of Chemistry; Georgia Institute of Technology, Atlanta, GA 30332, USA
[4]Department of Physics; University of South Florida, Tampa, FL 33620, USA

*Corresponding author. Email: faisal.alamgir@mse.gatech.edu

**Abstract:** Supermagnetic states are typically set during synthesis. Changing and retaining these states requires sustained external stimulus. Without resynthesis or continued stimulus, we show, for the first time, a reversible magneto-ionic switching between superferromagnetic (SFM) and superspin-glass (SSG) states using room-temperature Li-ion transport. We utilize the incomplete lithiation of NiO into nanoscale Ni and $Li_2O$ to switch between the states. Lithiation generates a fragmented nano-Ni network that strengthens dipolar interactions and promotes SFM order with stronger memory and higher room-temperature moments. Delithiation partially regenerates NiO and drives the system into an SSG state with higher low-temperature coercivity and sharper temperature-selective memory. The interconversion between SFM and SSG persists at all measured subsequent lithiation-delithiation cycles, creating a general framework for supermagnetic switching.

**Main Text:**

In magnetic nanoparticles, atomic spins combine into large moments known as superspins(*1*). The collective behavior of these spins gives rise to supermagnetism. Depending on their interactions, superspins can remain effectively independent, as in superparamagnetism (SPM), or develop interacting states such as in superspin glass (SSG) and in superferromagnet (SFM)(*2–4*). The interacting SSG and SFM states are particularly intriguing because they exhibit pronounced memory, aging and rejuvenation(*5*), offering functionality beyond simple modulation of magnetization or coercivity. Yet these states are typically set during synthesis(*6*, *7*) through a combination of particle size, composition and spacing, or accessed only under a sustained external stimulus(*8*) that temporarily modifies interaction. This limited accessibility constrains both fundamental study and their technological use. Here, we show reversible switching between interacting SFM and SSG states for the first time, without resynthesis or sustained external stimulation (**Fig. 1A**).

We achieve this using magneto-ionics, in which ion transport and chemical reactions instill contrasting non-volatile magnetic properties due to reconfigured physical and chemical states of a material. Magneto-ionics, however, has not yet been used for supermagnetic switching (*9*). We revisit $Li^+$ magneto-ionics(*10–12*) through the lens of lithium-ion battery and exploit three characteristic limitations of conversion electrodes: large compositional changes(*13*), in situ nanoparticle formation(*14–16*) and incomplete structural reversibility(*17*). Despite decades of study on conversion chemistry for energy storage (*18*, *19*), showing the intermixing between the reaction product, its magnetic evolution has generally been described in terms of ferromagnetic to and noninteracting SPM transitions(*20–24*). Here, we show for the first time that electrochemically generated nanoparticles can instead develop interacting supermagnetic states.

We use NiO as a model system because Ni and NiO sit at opposite ends of the magnetic spectrum: Ni is a ferromagnet, while NiO is an antiferromagnet. Although this contrast has made Ni/NiO heterostructures attractive for catalysis(*25*), spintronics(*26–28*) and magnetic switches(*29*, *30*), conventional interconversion relies on gas-phase oxidation or reduction(*31–34*), often at elevated temperature. $Li^+$ conversion provides a room-temperature, non-volatile route. Lithiation partially breaks NiO to form Ni nanoparticles dispersed within diamagnetic $Li_2O$, while delithiation reduces the Ni fraction and partially reforms NiO. Cyclic voltammetry (CV), X-ray diffraction (XRD), scanning transmission electron microscopy with electron energy-loss spectroscopy (STEM+EELS), and superconducting quantum interference device (SQUID) magnetometry, were used to track the chemical and magnetic reconstruction. In the discharged state, the high density of Ni nanoparticles strengthens interparticle coupling, while spatial interruption by NiO and $Li_2O$ prevents the formation of bulk ferromagnetism. Together, these conditions produce an SFM state. Delithiation dilutes the Ni network, increases magnetic heterogeneity and competing interactions, driving the system into an SSG state. Electrochemical conversion therefore reversibly switches not simply the magnetization, but the interaction landscape itself, transforming between SFM and SSG states within the same material.

**Conversion reaction reversibly reconstructs two nanoparticle heterostructures**

Electrochemical cycling of $Li^+$ ions into NiO follows the following one-step, reversible conversion reaction

$$2Li^+ + NiO \leftrightarrow Ni + Li_2O + 2e^-$$

as confirmed by the repeating single peak in the CV (**Fig. 1B, S1-2**). The first discharge (lithiation) peak sits at a lower potential (~0.2 V) due to the activation barrier and solid-electrolyte interphase (SEI) formation(*35*). The redox peaks become stable thereafter. However, XRD and SEM (**Fig. 1C and S3,4**) both show that this reaction reconstructs the initially homogeneous NiO electrode into a heterogeneous nanocomposite of metallic Ni embedded within an insulating matrix. This indicates partial conversion of Ni ↔ NiO, then a complete one(*19*). This small distinction becomes important in the magnetism studies in the future sections.

Structural evolution was monitored for the pristine electrode (P_NiO), the first three discharged states (NiO_1D, NiO_2D and NiO_3D) and the corresponding charged states (NiO_1C and NiO_2C) (**Fig. 1C-D**). While P_NiO is phase-pure (~36.4 nm crystallites), the discharged and charged states construct two distinct reproducible states: Ni/$Li_2O$/NiO (discharged) and Ni/NiO (charged). Upon cycling, metallic Ni nanoparticles nucleate and remain present throughout the cycles, slightly decreasing during charge. $Li_2O$ appears reversibly during discharge and disappears during charge. Both Ni and $Li_2O$ show coarsening around 3$^{rd}$ discharge(*24*). Rietveld refinement (**Fig. 1E**) shows that discharged electrodes are dominated by Ni (~60-70%), followed by equal parts NiO/$Li_2O$, while charged electrodes have a Ni(~40%)/NiO(~60%) mixture due to partial conversion of Ni ↔ NiO.

Electron microscopy independently confirms these architectures (**Fig. 1F and S5**). STEM-EELS mapping reveals spatially segregated Li-rich regions in the discharged case, consistent with the formation of a $Li_2O$-rich matrix surrounding metallic Ni NPs and residual NiO. Upon charging, the Li signal substantially weakens, while the remaining Li, O and Ni signals become more spatially overlapped, consistent with $Li_2O$ removal and partial reformation of NiO while increasing Ni/NiO interfaces.

Together, these results demonstrate that electrochemical cycling does not simply vary the amount of metallic Ni. Instead, it reversibly reconstructs two distinct nanoscale heterostructures, a Ni/$Li_2O$/NiO architecture during discharge and a conventional(*25*, *31*, *34*, *36*) Ni/NiO architecture during charge. These structural motifs define the exchange pathways and interparticle spacing between magnetic nanoparticles, providing the structural framework for the reversible switching between collective magnetic states described below.

### Discharged and charged states toggle between high and low magnetic moment at all measured temperatures

To determine how the reconstructed heterostructures modify the magnetic state, magnetization-field (M-H) loops were measured between ±70 kOe at 10K for pristine NiO and representative charged and discharged electrodes (**Fig. 2A,B**). All measurements were conducted on two independently prepared electrodes to test for reproducibility, with the magnetic-field axis corrected for flux-pinning effects using a Pd reference (**Fig. S6)**. The measurements were also conducted for 50, 300 and 325K (**Fig. S6-9**). Samples retain hysteresis and saturation magnetization ($M_s$) trend at all measured temperatures.

P_NiO exhibits the expected nearly linear antiferromagnetic response with negligible remanence and magnetization(*37*)(**Fig. 2A**). In contrast, all converted electrodes display nonlinear magnetization together with finite remanence and open hysteresis loops that persist to at least 325 K (**Fig. S7-9**), demonstrating the emergence of ferromagnetic-like order following electrochemical conversion. However, the 'S' shaped M-H curves in the charged case can be attributed to spin-glass systems(*5*) as the frustrated landscape prevents reaching $M_s$.

The persistence of hysteresis is unexpected given the ~2 nm Ni crystallites measured by XRD. Using the Néel-Brown model(*38*) with the bulk anisotropy of Ni(*39*) yields an isolated-particle blocking temperature, $T_B$ below 1 K (**Table S1**). Although the absolute $T_B$ is expected to vary because of nanoscale anisotropy and interface effects, isolated Ni nanoparticles cannot account for the above room temperature hysteresis. Instead, the magnetic response may be stabilized by collective interactions arising from the reconstructed nanoscale architecture(*40*).

The discharged electrodes consistently exhibit larger $M_s$ at all measured temperatures (**Fig. 2B and Fig. S9**) due to their Ni-rich heterostructure, whereas the charged Ni/NiO architecture produces substantially lower $M_s$ (*41*) despite retaining ~40% metallic Ni. This reproducible contrast demonstrates that electrochemical cycling reversibly switches between two magnetic states by reconstructing the underlying nanoscale heterostructure. At 10K the opposite trend is noticed for coercivity, $H_c$ as the charged electrodes show higher coercivity due to higher AFM NiO (~60%) compared to discharged cases (less than 30%).

While $M_s$ shows a reproducible trend for all measured temperatures, $H_c$ is variable above 10 K. We therefore quantify the $M_s$ as a metric for reversible magnetic switching, while $H_c$ above 10K is taken as qualitative measure indicating both charged and discharged heterostructures behave as soft magnets ($H_c$ <170 Oe).

The converted electrodes further exhibit squareness ($M_r/M_s$ = 0.28 - 0.37 at 10 K; **Table S2**) well below the ideal Stoner-Wohlfarth value of 0.5 expected for non-interacting blocked particles(*42*). Together with the high temperature hysteresis, these results exclude isolated superparamagnetism (SPM) as a potential magnetic state(*38*). Instead, electrochemical conversion generates interacting magnetic nanoparticle networks(*43*) whose collective behavior is reversibly tuned by the spatial arrangement of Ni, NiO and $Li_2O$. The nature of these collective states is resolved through temperature-field-dependence, susceptibility, memory and Henkel measurements below.

**Switching between SFM and SSG**

The hysteresis measurements indicate that electrochemical conversion generates interacting magnetic nanoparticles rather than isolated SPM. To probe the underlying collective behavior, zero-field-cooled (ZFC) and field-cooled (FC) magnetization were measured between 10 and 390 K under applied fields of 10, 100, 500, 3000 and 5000 Oe (**Fig. 2C and S10-17**).

At 500 Oe, all converted electrodes exhibit pronounced ZFC-FC bifurcation together with broad characteristic temperatures ($T^*$ = 104 - 170.1 K; **Fig. 2C and S13,14**). Because hysteresis persists above $T^*$ (to at least 325 K), these temperatures cannot represent conventional single-particle blocking $T_B$. Instead, they reflect the onset of collective freezing within the reconstructed nanoparticles assemblies.

The thermomagnetic response differs between the two electrochemical states. The charged Ni/NiO electrodes develop a plateau-like FC magnetization upon cooling due to frozen interacting superspins. This behavior is consistent with magnetic frustration arising in canonical-/cluster-spin(*44*) or superspin (SSG)-glasses(*45*). In contrast, the discharged Ni/NiO/$Li_2O$ electrodes continue to increase in FC magnetization, indicating that a larger fraction of moments stay field responsive. Together with their larger saturation magnetization, this behavior suggests a more strongly correlated ferromagnetic or a superferromagnetic (SFM)-like state(*46*). Furthermore, magnetic frustration limits superspin alignment in the SSG, making $T^*$ less field-dependent, whereas the SFM readily aligns its correlated domains, producing a stronger field response(*7*). Increasing the measuring field from 500 to 3000 and 5000 Oe (**Fig. S17**), the discharged NiO_2D decreases the $T^*$ (<10K) consistent with SFM-like behavior. Whereas the decrease is less prominent in charged NiO_2C (51.5 and 21.7 K at 3k and 5kOe, respectively) consistent with SSG.

**Charged state exhibits slow dynamics**

DC magnetometry establishes the presence of supermagnetism but does not distinguish between the slow or fast dynamic of superspins. A frustrated SSG should have slower dynamics compared to an SFM-like system. We therefore measured temperature-dependent AC susceptibility, where the in-phase component ($\chi'$) probes the reversible magnetic response and the out-of-phase component ($\chi''$) measures dissipative relaxation.

At 100 and 1000 Hz (**Fig. 3A and S18,19**), $\chi'$ increases with temperature for all converted electrodes, with the discharged NiO_2D and NiO_3D electrodes exhibiting the largest susceptibility, consistent with their higher saturation magnetization. In contrast, the charged NiO_1C and NiO_2C electrodes display broad $\chi''$ features over an extended temperature range, demonstrating slow magnetic relaxation arising from a broad distribution of interaction. Such broad dissipation is characteristic of heterogeneous interacting nanoparticle systems, including SSG(*47*).

Frequency-dependent measurements further emphasize the contrast between the two electrochemical states (**Fig. 3B**). The discharged NiO_2D electrode exhibits smooth $\chi'$ and $\chi''$ responses without a well-defined frequency-dependent maximum between 1 and 100 Hz, indicating that its dominant relaxation processes are either fast making it SFM-like(*48*) or lie outside the experimental window. By contrast, the charged NiO_2C electrode develops broad, frequency-dependent features near 80 - 130 K in both $\chi'$ and $\chi''$, consistent with cooperative slowing of interacting superspins. The weak frequency dependence of $\chi'$ further supports a broad distribution of relaxation times.

Overall, the charged Ni/NiO heterostructure exhibits a slow relaxation characteristic of an SSG-like state, whereas the discharged Ni/$Li_2O$/NiO architecture has a faster response in the same experimental set-up. The distinction between these two regimes is further established through magnetic memory and Henkel measurements below.

**Discharged state exhibit stronger memory while charged state show sharper temperature-selective memory**

Magnetic memory directly probes nonequilibrium dynamics by testing whether a system retains information about its thermal history. In an SSG state, below ($T_f$), superspins occupy a rugged free-energy landscape and progressively relax into deeper metastable minima during the waiting time, $t_{wait}$ (**Fig. 3C**). Upon reheating, the system encounters the configurations established during the stop, producing a memory feature near the corresponding waiting temperature.

To further distinguish the collective magnetic states, ZFC and FC memory measurements were performed with 3 h stops at 30 and 60 K (**Fig. 3D and S20**). The discharged electrode exhibits broad, weakly temperature-selective deviations from the reference curves (**Fig. 3D**), consistent with a correlated SFM state(*46*) in which memory is distributed through the pinning and reorganization of ferromagnetic-like domains. In contrast, the charged electrode shows more localized peaks near the waiting temperatures, most prominently at 60 K as low-temperature peaks have been shown to be missing in multi-stop memory measurement in SSG due to low resolution(*44*). Such temperature-selective memory is characteristic of glassy freezing.

Interestingly, while charged states show sharper temperature-selective memory, the discharged state shows a ~2.7 times stronger memory, which makes accessing different supermagnetic states a lucrative prospect for neuromorphic application (*49*).
Together, the broad memory of the discharged state and more temperature-selective memory of the charged state support SFM- and SSG-like collective behavior, respectively rather than the existence of SPM state(*3*).

**Henkel analysis reveals stronger interactions in discharged state**

Henkel analysis further reveals distinct interaction regimes between the discharged and charged states at 10K (**Fig. 4A and S21**). NiO_2D exhibits a pronounced overshoot ($m_R$, $m_D > 1$) above unity, whereas NiO_2C evolves more gradually and displays a substantially higher coercive field (**Fig. S21a**). The overshoot in NiO_2D may reflect the training behavior also observed in its M-H loops. The extracted coercivities(*50*) are consistent with those obtained independently from the 10 K M-H loops.
The corresponding $\Delta M$(*51*), curves are predominantly negative, indicating that demagnetizing interactions initially oppose collective reversal, generally attributed to dipole-dipole interaction favoring the demagnetized state(*52*). While initially NiO_2D shows stronger dipolar interaction, it quickly approaches and crosses $\Delta M = 0$ at low field, developing a small positive contribution(*50*) at higher fields (**Fig. 4A**). This indicates at high field, superspins align readily to the applied field overcoming its inherent frustrated landscape due to larger ferromagnetic Ni. This positive deviation is often attributed to exchange interactions(*50*). Whereas the negative $\Delta M$ persists for NiO_2C making it an SSG-like state(*53*).
Moreover, the $m_D$-$m_R$ curves show a stronger deviation from the Stoner-Wohlfarth line for NiO_2D (**Fig. 4A and S21b**) indicating a strong interacting system(*54*, *55*) compared to NiO_2C. This stronger interaction compared to SSG-like NiO_2C might rule out the possibility of NiO_2D being an interacting SPM (I-SPM) system(*56*). Together, the training behavior and earlier recovery of $\Delta M$ indicate a more strongly correlated reversal process in the discharged state, consistent with its SFM-like collective behavior.

**Conversion reaction modulates both the concentration and the magnetic interaction landscape**

To directly visualize how electrochemical conversion reorganizes the magnetic interaction network, the nanoscale phase distribution was examined by STEM and STEM-EELS (**Fig. 4B**). False-color STEM images constructed using the Ni, NiO and $Li_2O$ reflections identified from the corresponding FFTs (**Fig. S22**) reveal extensive nanoscale intermixing of the conversion products in both states. Consistently, DF-STEM and EELS mapping distinguish bright NiO-rich and dark Ni-rich regions.
Importantly, the magnetic state evolves systematically with composition (**Fig. 4C**). Ni-rich discharged electrodes, with lower NiO fractions and higher saturation magnetization, fall within the SFM regime, whereas electrochemical reoxidation increases the NiO fraction, suppresses the net magnetization and drives the system into the SSG regime. The transition near ~30 mole% NiO therefore suggests that electrochemical cycling controls not only the amount of ferromagnetic Ni, but also the connectivity and interaction landscape of the nanoscale magnetic network, enabling reversible switching between SFM- and SSG-like collective states.

**Conclusion**

Reversible reconfiguration between distinct supermagnetic states has not been demonstrated prior to this work, partly because substantial structural and compositional reorganization is required to transform a strongly coupled SFM state to a frustrated SSG state. The expectation, therefore, is for such a transformation to be irreversible. We bypass this constraint by taking advantage of conversion electrode chemistry that, for energy storage purposes, must undergo large compositional reconstruction reversibly and over thousands of electrochemical cycles. Furthermore, regardless of the initial particle size, subsequent conversion drives the electrode into the supermagnetic size regime because kinetically limited nanoparticle formation is inherent to the reaction(*14–16*). Despite recent magnetism studies of conversion electrodes(*20*), their collective supermagnetic behavior has never been discovered. In this work, we show that electrochemically generated conversion products exhibit magnetic memory, revealing collective magnetic behavior. This is also the first systematic study of magnetism in NiO conversion electrode. This is a notable oversight given that NiO is a widely studied correlated insulator used across optoelectronic, electronic and magnetic devices. Magnetism studies in Li-ion conversion electrodes have largely focused on Fe-based systems(*12*, *20–22*), leaving the collective magnetic behavior of NiO comparatively unexplored.

The principle of generating magnetic nanoparticles within conversion electrodes for switching between collective supermagnetic states could be extended beyond NiO to other systems. Furthermore, spatial mixing of nanoparticles, which dictates the distinct supermagnetic state, can be controlled by the electrochemical reaction rate(*15*). The collective supermagnetic states may have been missed by earlier studies(*12*, *20–22*, *24*) due to improper spatial mixing of conversion products. Greater structural control through thin films, porous architectures or size-controlled nanoparticles could provide finer control over these interactions(*19*, *35*, *57–62*). Conversion chemistry therefore offers both a route to programmable supermagnetism and a platform for constructing and investigating interacting superspin networks far from equilibrium.

**Acknowledgments:**

**Funding:**

Georgia AI Manufacturing (GA-AIM) project funded through Economic Development Agency

Air Force Office of Scientific Research (AFOSR) under the support provided by the Organic Materials Chemistry Program (Grant FA9550-23-1-0654, Program Manager: Dr. Michael Berman)

**Author contributions:**

Conceptualization: SA

Supervision: FMA

Investigation: SA, YJY, AHC

Visualization: SA, FMA, YJY, JW, AHC, DAA

Funding acquisition: FMA, JDA

Writing – original draft: SA

Writing – review & editing: FMA, DAA, AHC, TAB

**Competing interests:** Authors declare that they have no competing interests.

**Data, code, and materials availability:** All data are available in the main text or the supplementary materials.

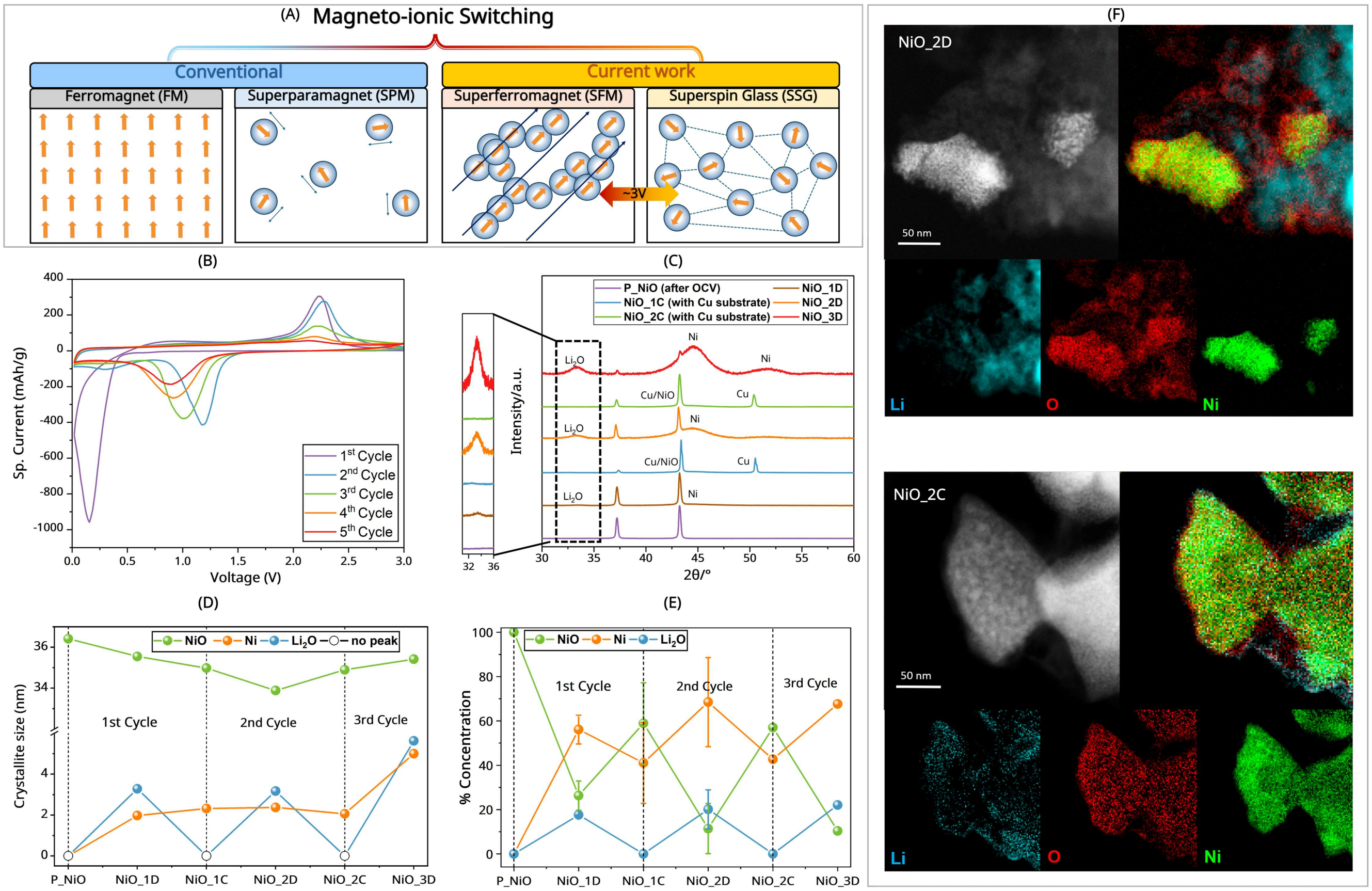

**Fig. 1. Magneto-ionic reconstruction of collective supermagnetic states in NiO.**
**(A)** Schematic comparison of conventional magneto-ionic switching between ferromagnetic (FM) and superparamagnetic (SPM) states with the reversible switching between superferromagnetic (SFM) and superspin-glass (SSG) states demonstrated here.
**(B)** Cyclic voltammetry of NiO in a $LiPF_6$-based half-cell (0.02-3.0 V vs. $Li/Li^+$, 0.1 mV $s^{-1}$) for the first five electrochemical cycles, showing characteristic activation and reversible conversion behavior.
**(C)** XRD of P_NiO and electrodes after successive discharge (D) and charge (C) steps, showing reversible formation and partial oxidation of metallic Ni and $Li_2O$.
**(D)** Evolution of the crystallite size of NiO, Ni and $Li_2O$ extracted from XRD peak broadening.
**(E)** Phase fractions of NiO, Ni and $Li_2O$ obtained from Rietveld refinement over three electrochemical cycles. Data represents the mean of two independent electrodes; bars indicate the range.
**(F)** STEM-EELS elemental maps of representative NiO_2D and NiO_2C regions showing the spatial distributions of Li (cyan), O (red) and Ni (green) and their corresponding composite maps. Scale bars, 50 nm.

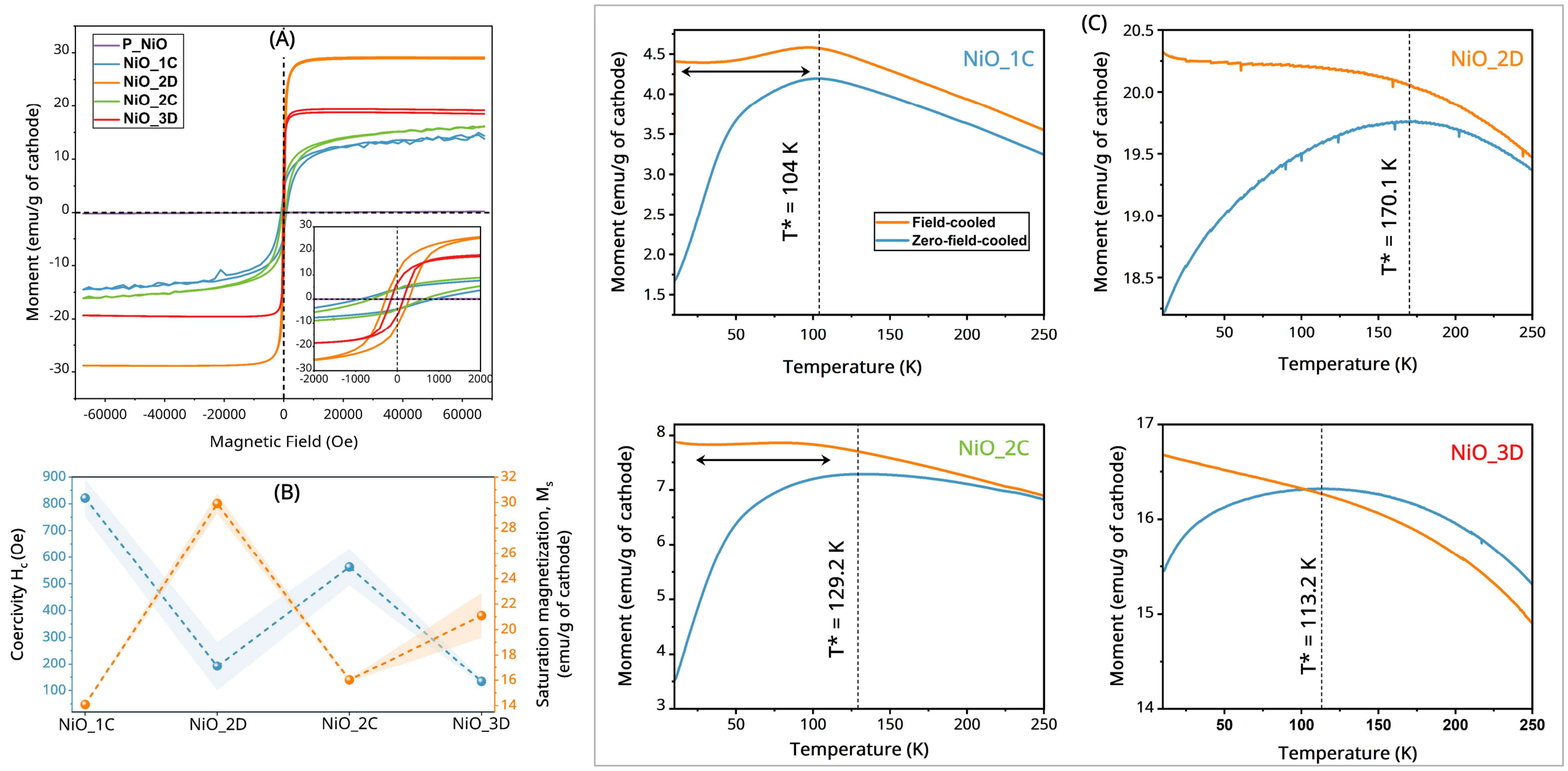

**Fig. 2.**
**Magnetic state evolution across electrochemical states of NiO.**
**(A)** Magnetic hysteresis loops of P_NiO and electrochemically cycled NiO states measured at 10K between ±70 kOe, with the low-field (±2 kOe) region enlarged in the inset.
**(B)** Coercivity, $H_c$, and saturation magnetization, $M_s$, extracted from the hysteresis loops for successive charged and discharged states at 10K. Shaded regions represent the range obtained from two independently prepared electrodes for each electrochemical state.
**(C)** ZFC and FC magnetization measured between 10-250 K under an applied field of 500 Oe for NiO_1C, NiO_2D, NiO_2C and NiO_3D. Dashed vertical lines mark the characteristic temperatures, $T^*$.

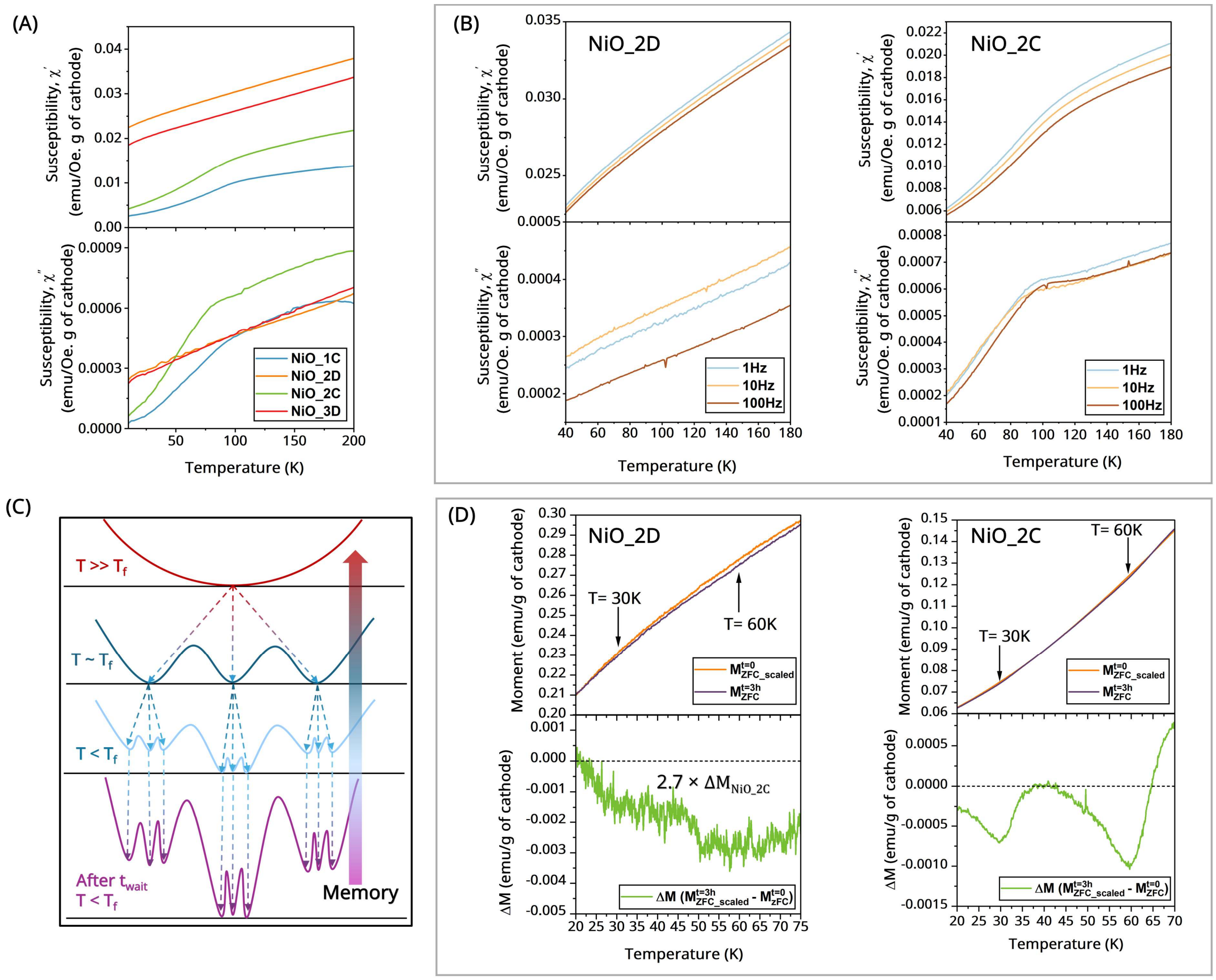

**Fig. 3. Dynamic susceptibility and magnetic memory distinguish supermagnetic states.**
**(A)** Temperature dependence of the real ($\chi'$) and imaginary ($\chi''$) components of the AC susceptibility for charged and discharged NiO states measured at 100 Hz using 10 Oe AC excitation and DC bias field.
**(B)** Frequency-dependent $\chi'$ and $\chi''$ for NiO_2D and NiO_2C measured at 1, 10 and 100 Hz using a 2 Oe AC excitation field with zero DC bias field.
**(C)** Schematic energy-landscape representation of memory formation as the system is cooled through the freezing temperature, $T_f$, and aged at an intermediate temperature for $t_{wait}$.
**(D)** ZFC memory measurements for NiO_2D and NiO_2C using 3-hour cooling stops at 30 and 60 K. Reference curves acquired without stops are compared with the stop-and-wait curves measured on reheating under 10 Oe. Difference curves, ΔM, isolate the memory contribution associated with the temperature stops. For visualization of the memory response, the t = 0 reference curves were scaled to the corresponding t = 3h curves, such that ΔM corresponds to a zero baseline. The unscaled curves are shown in **Figure S20**.

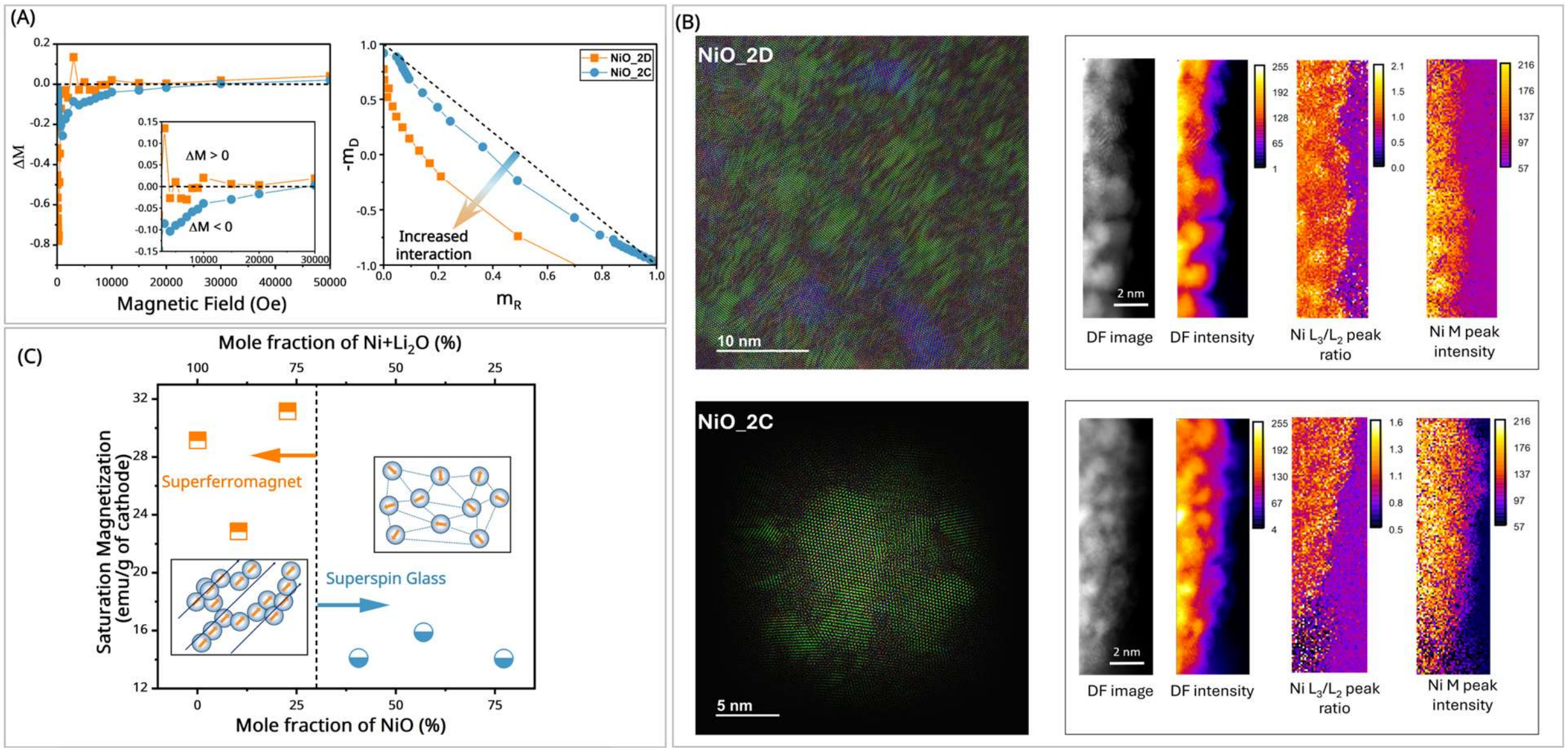

**Fig. 4. Interparticle interactions and nanoscale reconstruction underpin switching between superferromagnetic and superspin-glass states.**

**(A)** Henkel analysis of NiO_2D and NiO_2C reveals distinct interaction signatures. The Henkel parameter, ΔM, as a function of applied field (left) shows distinct interaction signatures in the discharged and charged states; the inset highlights the sign of ΔM. Normalized demagnetization remanence, $-m_D$, plotted against isothermal remanent magnetization, $m_R$ (right), is compared with the noninteracting Wohlfarth relation (dashed line), with deviations indicating increased interparticle interactions.

**(B)** False-color STEM images (left) of NiO_2D and NiO_2C show the distribution of NiO (green), Ni (red) and $Li_2O$ (blue), assigned from the corresponding real-space images and FFTs (**Fig. S22)**. Dark-field (DF) STEM images and corresponding DF intensity, Ni $L_3/L_2$ peak-ratio and Ni M-edge intensity maps (right) resolve the local Ni distribution and oxidation state across the interface. The Ni $L_3/L_2$ white-line ratio distinguishes metallic Ni (darker region) from NiO (brighter region), while the Ni M-edge intensity maps identify the spatial distribution of Ni-containing regions across the interface. NiO_2D and NiO_2C both show an intermixed Ni/NiO within the Ni-rich domains.

**(C)** Saturation magnetization as a function of NiO and Ni + $Li_2O$ phase fraction shows the evolution from the Ni-rich SFM state toward the more oxidized SSG state. Schematics illustrate the corresponding superspin configurations.

# Supplementary Materials for

## Ionic switching between supermagnetic states

Sumaiyatul Ahsan[1], Yin-Ju Yen[2], Andrew H. Comstock[3], Tanya A. Balandin[1], Jason D. Azoulay[1,3], Jamie Warner[2], Dario A. Arena[4], Faisal M. Alamgir[1]*

### Materials and Methods

Electrode and coin-cell preparation

NiO electrode slurries were prepared by mixing NiO, carbon black and poly(vinylidene fluoride) (PVDF) in a 90:5:5 weight ratio, using N-methyl-2-pyrrolidone (NMP) as the solvent. The resulting slurry was coated onto a Cu current collector with a nominal wet-film thickness of 75 µm. NiO electrodes were assembled into coin-cell using Li metal as both the counter and reference electrode and $LiPF_6$ as electrolyte. Electrochemical cycling was performed between 0.02 and 3.0 V versus $Li/Li^+$. Cyclic voltammetry (CV) was conducted at 0.1 mV $s^{-1}$, unless otherwise specified. Electrodes were recovered at selected charged and discharged states, washed with dimethyl carbonate (DMC), and subsequently characterized by structural and magnetic measurements.

Sample-state nomenclature

Pristine NiO is denoted P_NiO. Electrochemically converted samples are identified by the cycle number followed by their electrochemical state, where C and D denote charged and discharged states, respectively. Thus, NiO_1C corresponds to the electrode recovered after the first charge, NiO_2D and NiO_2C to the second discharge and charge, respectively, and NiO_3D to the third discharge.

Electron microscopy

Electrode morphology was examined by scanning electron microscopy (SEM) under high-vacuum conditions. Scanning transmission electron microscopy (STEM) images were acquired with a JEOL NEOARM transmission electron microscope and an acceleration voltage of 200 kV, equipped with a Gatan Quantum Electron Energy filter and a 2D HPAD array direct electron detector (Quantum Detectors Medipix 1024×256) for low dose EELS. The EELS maps were processed with Gatan Digital Micrograph for background subtraction and elemental signal collections. The Ni $L_3/L_2$ intensity map was calculated with ImageJ by integrating peak areas of Ni-$L_3$ edge within 847.7 - 854.8 eV and Ni-$L_2$ edge within 865.0 - 873.1 eV. The Ni-$M_{2,3}$ edge intensity map was subtracted from 64.0 - 86.5 eV. The EELS map coloring was produced with ImageJ.

X-ray diffraction and crystallite-size analysis

Structural evolution following electrochemical conversion was examined by X-ray diffraction (XRD). Diffraction features were assigned to NiO, Ni and $Li_2O$, with contributions from the Cu current collector identified separately. Characteristic reflections were used to estimate the crystallite sizes of the constituent phases using Scherrer analysis.

DC magnetometry

Magnetic measurements were performed using SQUID magnetometry. Field-dependent magnetization was measured between ±70 kOe at temperatures from 10 to 390 K. Measurements were repeated using independently prepared electrodes to evaluate the reproducibility of the electrochemical-state-dependent magnetic response. Coercivity and saturation magnetization were extracted from the corresponding hysteresis measurements.

ZFC/FC magnetometry and magnet degaussing

Temperature-dependent magnetization was measured using zero-field-cooled (ZFC) and field-cooled (FC) protocols, primarily under applied fields of 10 to 5000 Oe. Because low-field measurements are particularly sensitive to residual magnetic fields in the superconducting magnet, the magnet was degaussed/demagnetized at 300 K by oscillating the applied field between ±30 kOe before the low-field measurements. Control measurements performed with and without this degaussing procedure (**Fig. S10,11**) were used to distinguish intrinsic sample responses from artifacts arising from residual trapped field in the superconducting magnet.

AC magnetic susceptibility

Temperature-dependent AC susceptibility was measured to probe the dynamic magnetic response of the electrochemically converted electrodes. The real, $\chi'$ and imaginary, $\chi''$, components of susceptibility and the corresponding phase angle were recorded under an AC excitation field of 2 and 10 Oe at 1, 10, 100 and 1000 Hz under DC bias fields of 0, 10 and 100 Oe.

Magnetic memory measurements

Magnetic memory was investigated using ZFC and FC stop-and-wait protocols under a 10 Oe field, with intermediate temperature stops at 60 and 30 K during cooling at 2K $min^{-1}$ rate to 10K. For the ZFC protocol, the field remained off during cooling and both temperature stops, whereas for the FC protocol, the 10 Oe field remained applied throughout cooling and the stops. Reference ZFC and FC curves were measured under the corresponding conditions without temperature stops. Difference curves, ΔM, were calculated to isolate the memory contribution.

## Supplementary Text

Pd-reference correction for flux pinning in the magnetic hysteresis

A palladium (Pd) reference was used to quantify the absolute magnetic-field error associated with remanent or trapped flux in the superconducting magnet. Because Pd exhibits a linear paramagnetic response at room temperature, the actual magnetic field was calculated from the measured Pd moment according to

$$H_{actual} = \frac{\mu_{Pd}}{\chi_g m_{Pd}}$$

where $\chi_g = 5.25 \times 10^{-6}$ emu $Oe^{-1}$ $g^{-1}$ is the mass susceptibility of Pd at 298 K and $m_{Pd}$ = 0.2684 g is the mass of the reference. The resulting difference between the calculated and nominal fields,

$$\Delta H = H_{actual} - H_{nominal}$$

was used to correct the magnetic-field axis of the NiO hysteresis data (**Fig. 2A and S6-8**). The magnetic moment of the electrode was not modified by this correction.

IRM/DCD and Henkel analysis

Interparticle magnetic interactions were evaluated using isothermal remanent magnetization (IRM) and DC demagnetization (DCD) measurements. For the IRM protocol, the sample was first brought to a demagnetized state (H = 0) by following the outlined degaussing protocol. Progressively increasing positive magnetic fields, H (up to 50 kOe), were then applied, with the field returning to zero after each step and the remanent magnetization $M_R$ (H) was recorded. For the DCD protocol, the sample was first saturated in the negative field direction(*51*) at -50 kOe. The field was then returned to zero, followed by progressively increasing positive reversal fields, H, each of which was removed before recording the corresponding raw remanent magnetization, $M_D$ (H).

The normalized IRM and DCD remanences were defined as

$$m_R(H) = \frac{M_R(H)}{M_R(\infty)}$$

And,

$$m_D(H) = \frac{M_D(H)}{M_R(\infty)}$$

where is $M_R$ (∞) the saturation IRM.

For an assembly of non-interacting, randomly oriented single-domain particles, the remanences follow the Wohlfarth relation

$$m_D(H) = 1 - 2m_R(H)$$

However, as we started with a negative saturation magnetization for DCD protocol(*51*), the Wohlfarth relation will be modified to following:

$$-m_D(H) = 1 - 2m_R(H)$$

For the same reason **Fig. 4A**, the Henkel plot was plotted for $-m_D$ vs $m_R$.

Hence, the deviations from the non-interacting reference were quantified using the Henkel parameter(*51*)

$$\Delta M(H) = 2m_R(H) - m_D(H) - 1$$

Accordingly, ΔM = 0 corresponds to the non-interacting Wohlfarth limit, while positive or negative deviations indicate that the measured DCD remanence lies above or below the corresponding non-interacting reference, respectively.

Theoretical Blocking Temperature:

The persistence of hysteresis to temperatures well above that expected for isolated ~2 nm Ni particles are particularly notable. The blocking temperature of a non-interacting, single-domain particle is (*38*)

$$T_B \approx \frac{KV}{25k_B}$$

where K is the effective magnetic anisotropy and V is the particle volume. Using the bulk Ni(*39*) anisotropy, $K \approx 5 \times 10^3$ J/m$^3$ and the crystallite dimensions obtained from XRD yields an isolated-particle blocking temperature below 1 K (**Table S1)**.

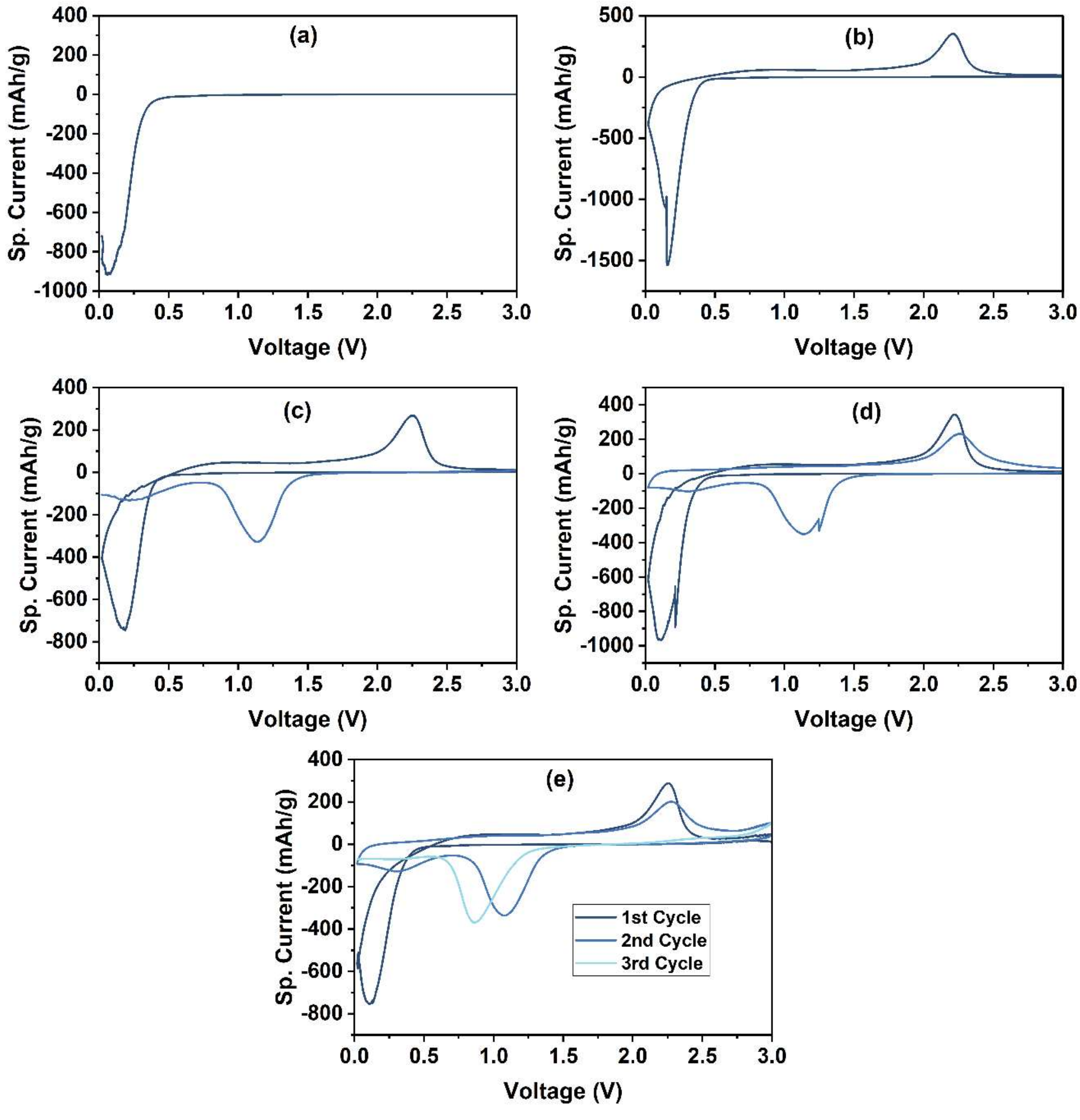

**Fig. S1. Electrochemical cycling of the NiO electrodes used for structural and magnetic characterization.**

CV profiles corresponding to the individual electrodes subsequently recovered: **(a)** NiO_1D, **(b)** NiO_1C, **(c)** NiO_2D, **(d)** NiO_2C and **(e)** NiO_3D. Measurements were performed between 0.02 and 3.0 V versus $Li/Li^+$ at a scan rate of 0.1 mV $s^{-1}$.

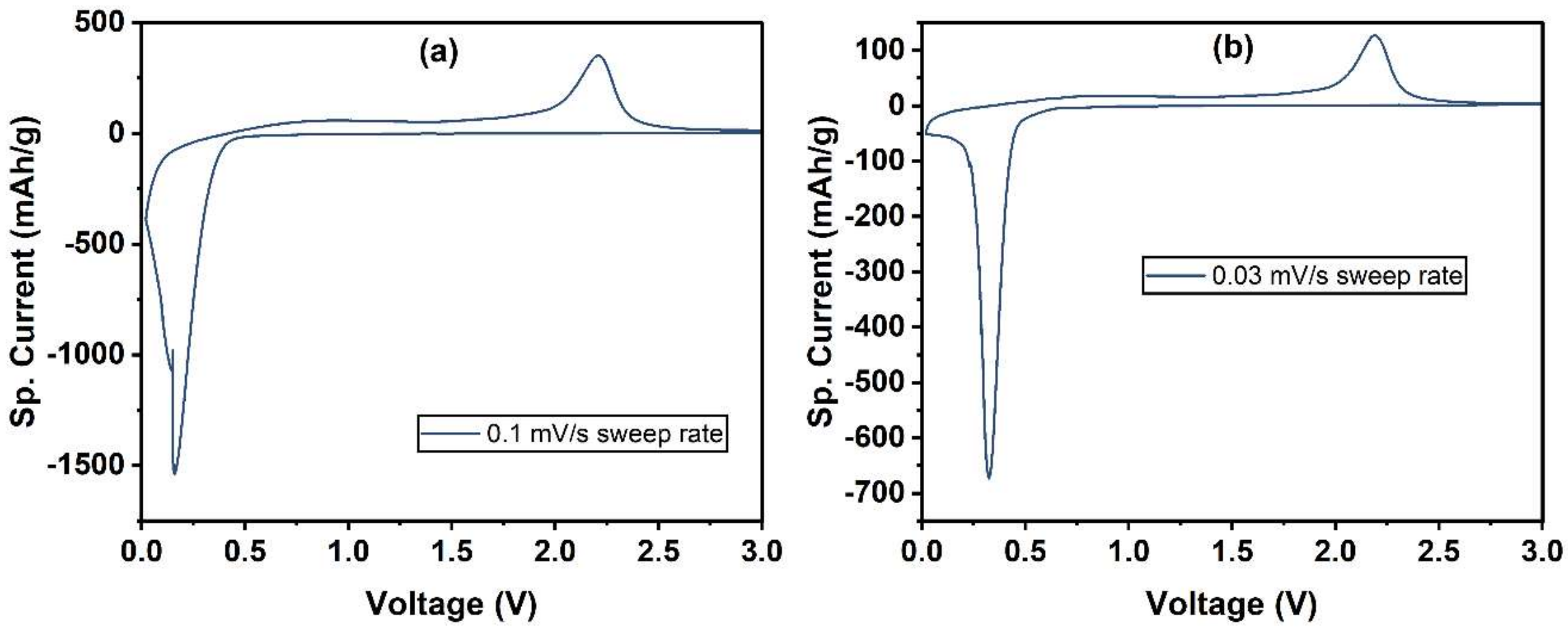

**Fig. S2. Effect of CV sweep rate on the electrochemical conversion of NiO.**

CV profiles of NiO electrodes measured between 0.02 and 3.0 V versus $Li/Li^+$ at scan rates of **(a)** 0.1 mV $s^{-1}$ and **(b)** 0.03 mV $s^{-1}$. Decreasing the sweep rate shifts the first-discharge reduction feature to higher voltage, consistent greater equilibration during the conversion reaction.

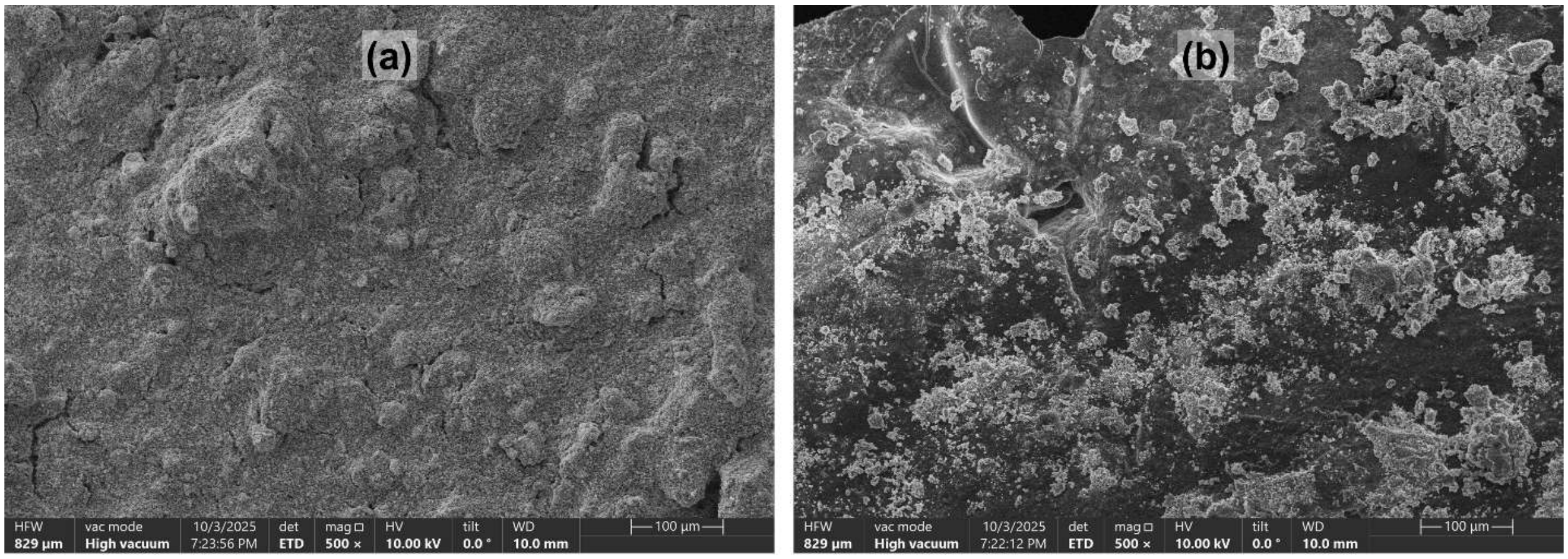

**Fig. S3. Morphological evolution of the NiO electrode during electrochemical conversion.** SEM images of the **(a)** P_NiO and **(b)** NiO_3D. Images were acquired under high-vacuum conditions at an accelerating voltage of 10 kV. NiO_3D exhibits a more heterogeneous and particulate surface compared with P_NiO.

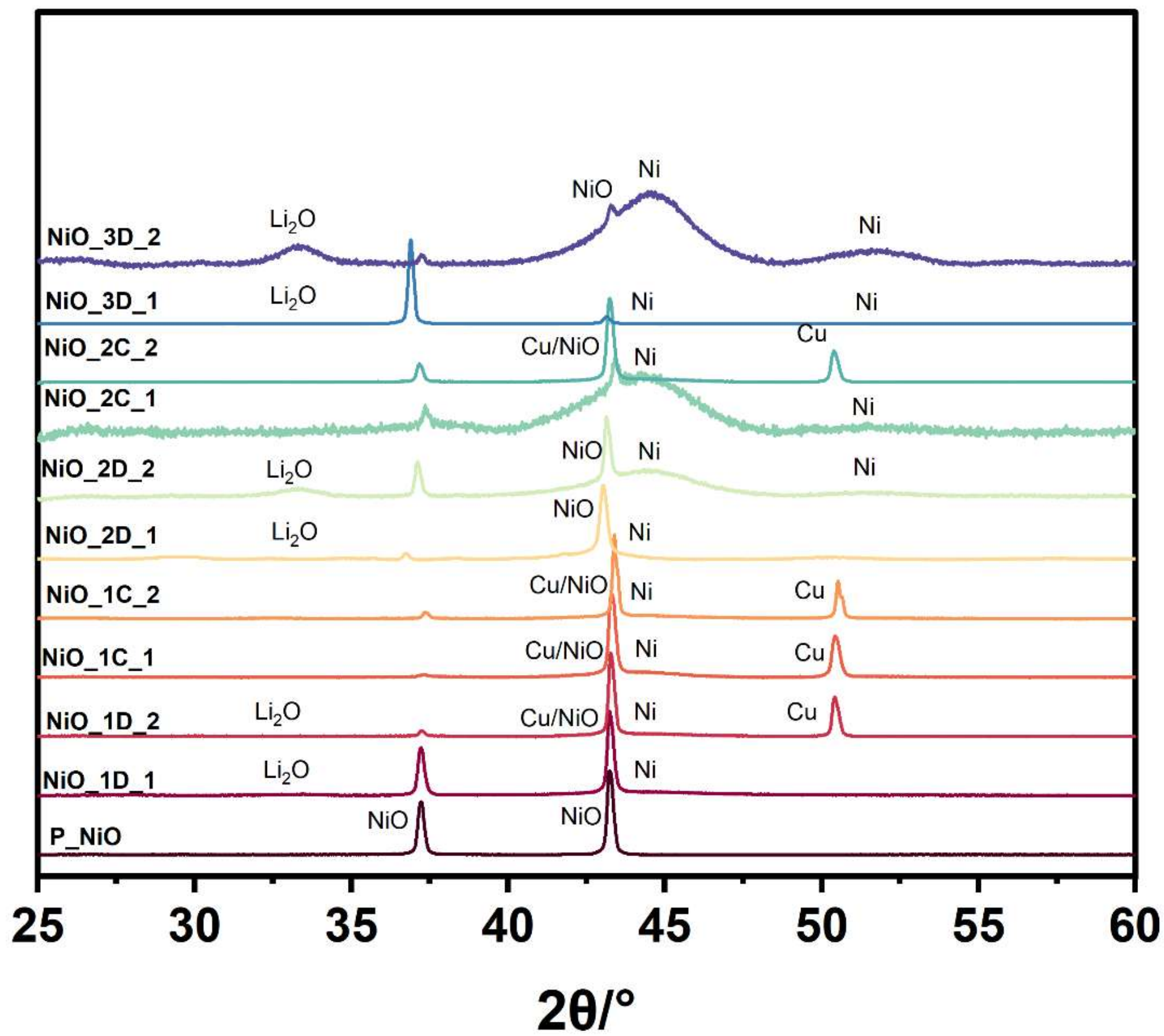

**Fig. S4. Structural evolution of NiO during electrochemical cycling.**

XRD patterns of pristine P_NiO and electrochemically converted electrodes recovered at different charged (C) and discharged (D) states. Characteristic reflections associated with NiO, Ni and $Li_2O$ are indicated. Reflections associated with the Cu current collector and overlapping Cu/NiO contributions are also labelled. NiO_3D_1 has large contribution from the sample holder due to low sample recovery. Two independently prepared electrodes are shown for each electrochemical state.

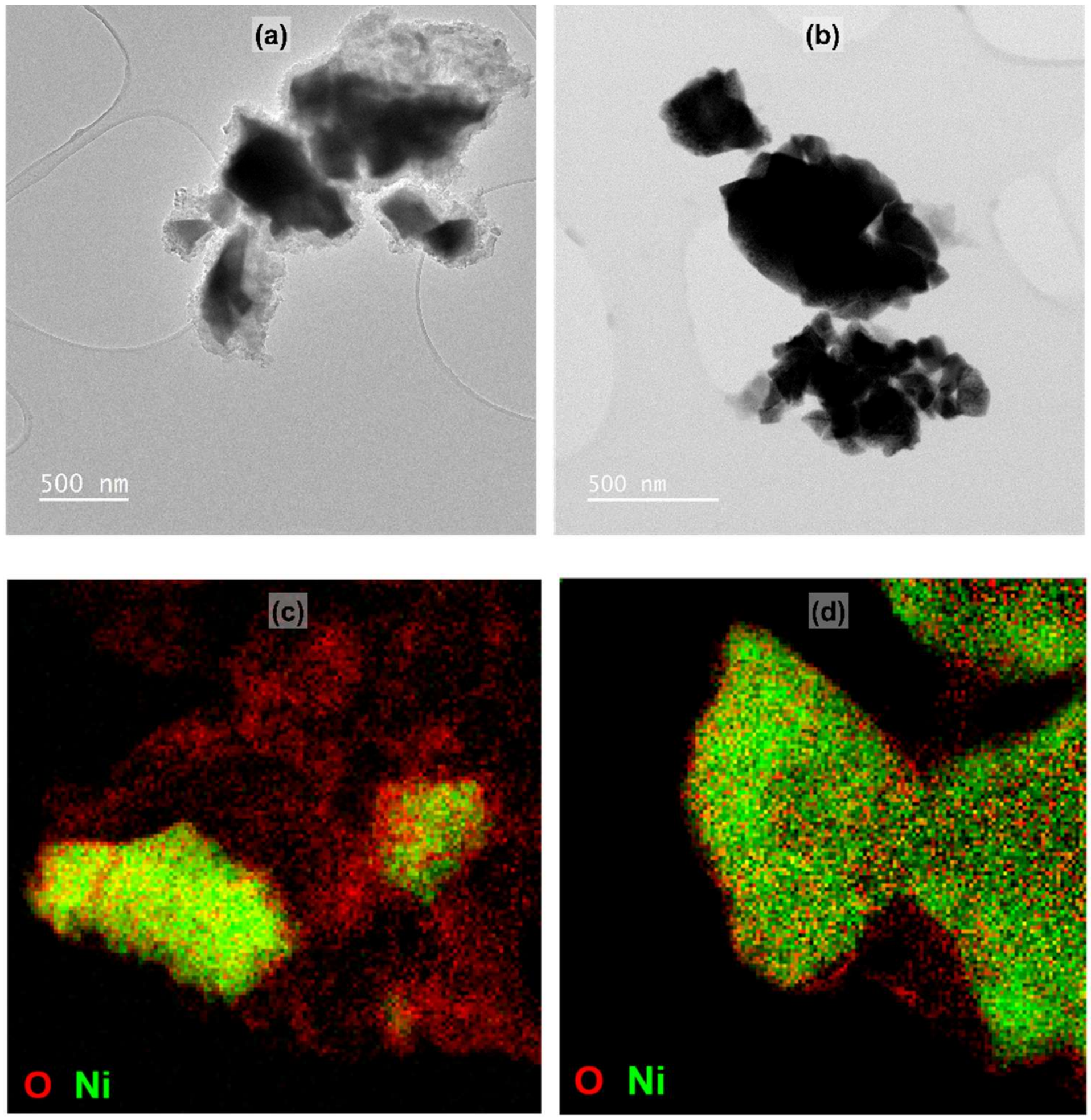

**Fig. S5. Nanoscale redistribution of Ni and O following electrochemical conversion in NiO_2D (a,c) and NiO_2C (b,d).**

**(a,b)** STEM-BF images showing larger Li-rich matrix in NiO_2D and Ni-rich regions in NiO_2C.
**(c,d)** corresponding elemental maps showing the spatial distributions of O (red) and Ni (green) in electrochemically converted NiO electrodes. Spatial overlap between the Ni and O signals corresponds to NiO, whereas O-rich regions lacking a corresponding Ni signal indicate oxygen-containing phases ($Li_2O$) spatially separated from Ni-rich regions.

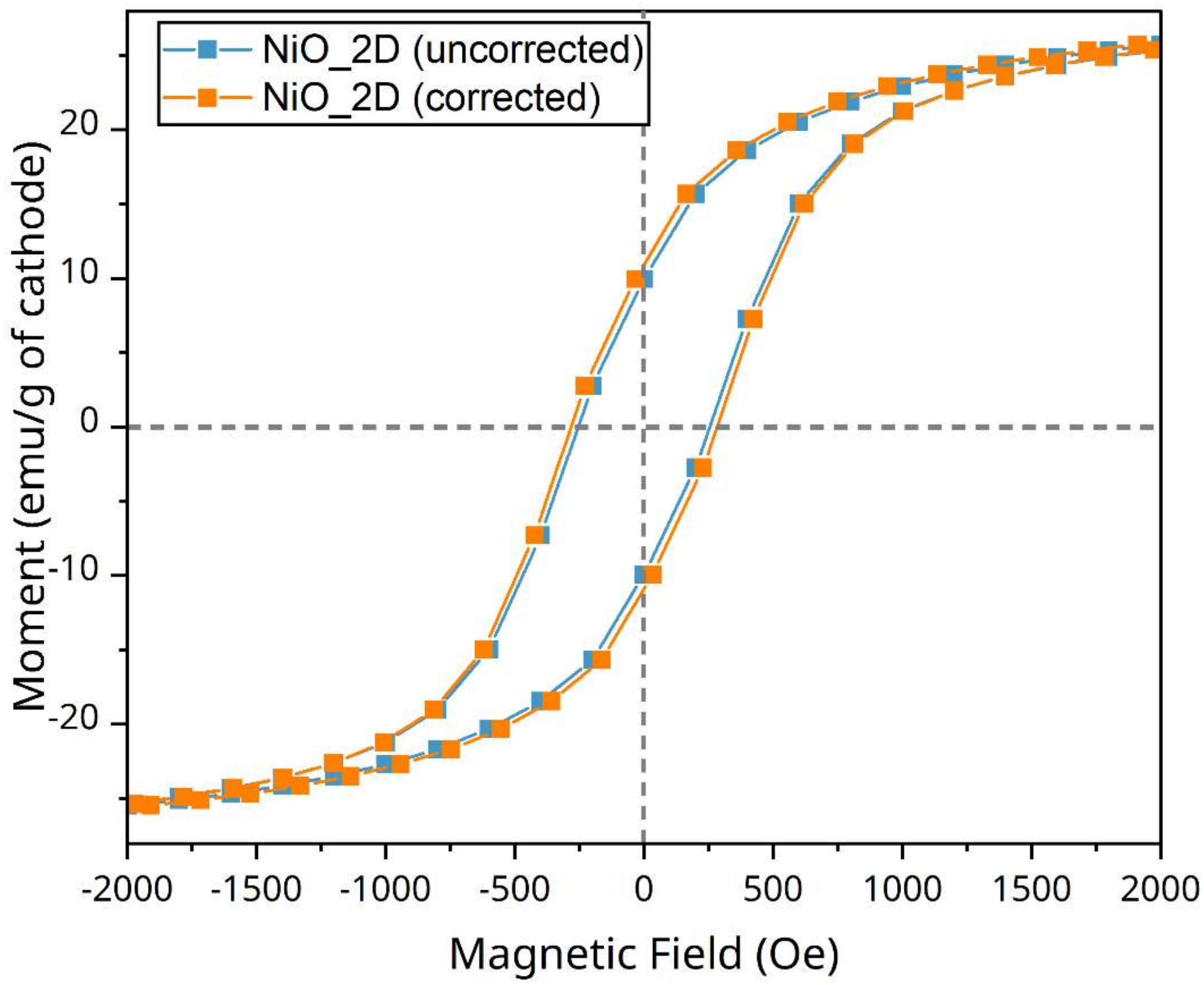

**Fig. S6. Pd-reference correction of the magnetic-field axis for flux pinning.**

Hysteresis loop of NiO_2D before and after correction of the nominal magnetic field using a Pd reference. The correction shifts the data along the field axis without altering the measured electrode moment. The close agreement between the corrected and uncorrected loops demonstrates that the observed hysteresis is retained after accounting for flux-pinning-induced field offsets.

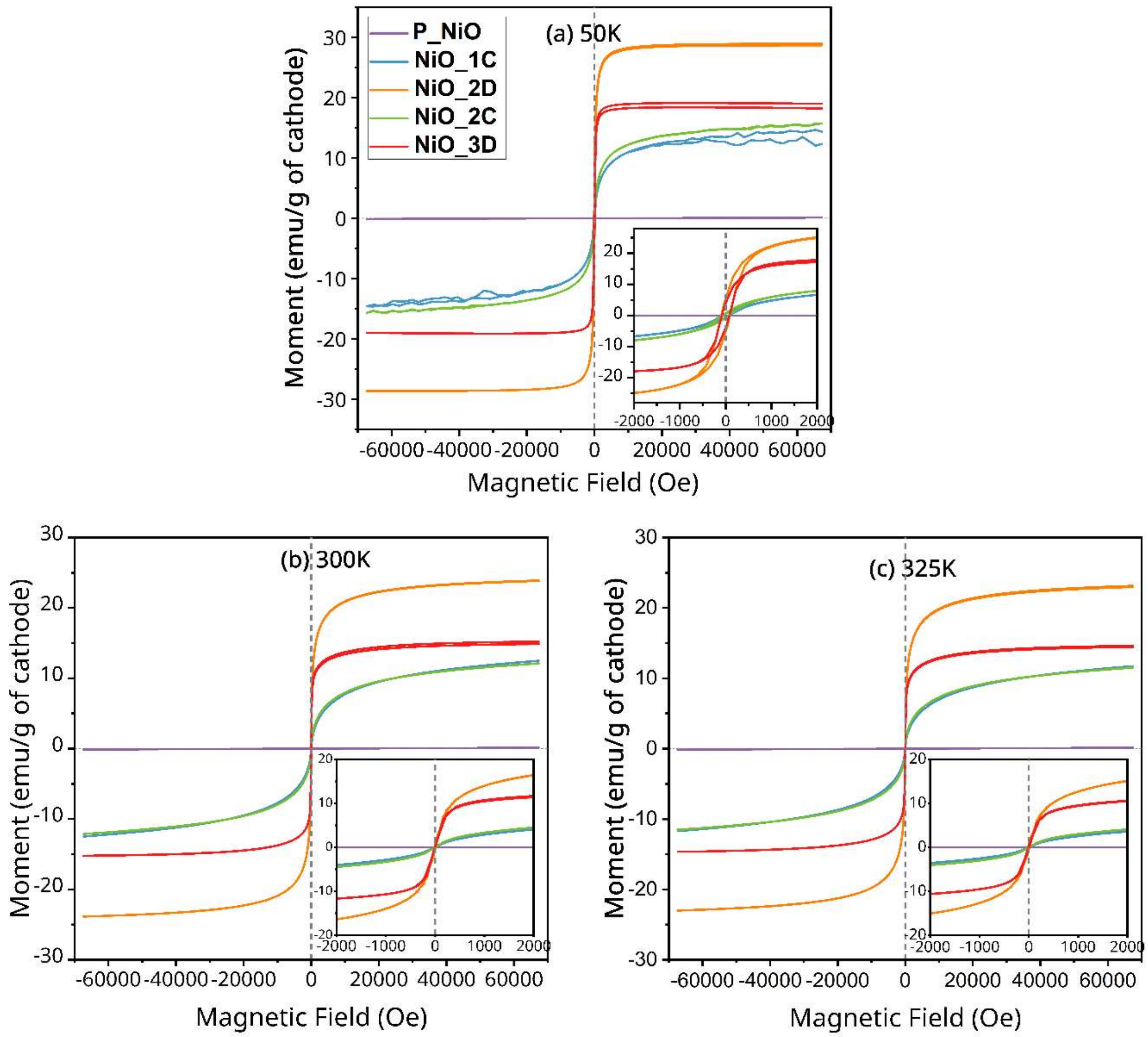

**Fig. S7. Magnetic hysteresis across electrochemically converted NiO states.**

Magnetic moment as a function of applied magnetic field for P_NiO and the NiO_1C, NiO_2D, NiO_2C and NiO_3D electrodes measured at **(a)** 50 K, **(b)** 300 K and **(c)** 325 K between ±70 kOe. Insets show enlarged views of the low-field region.

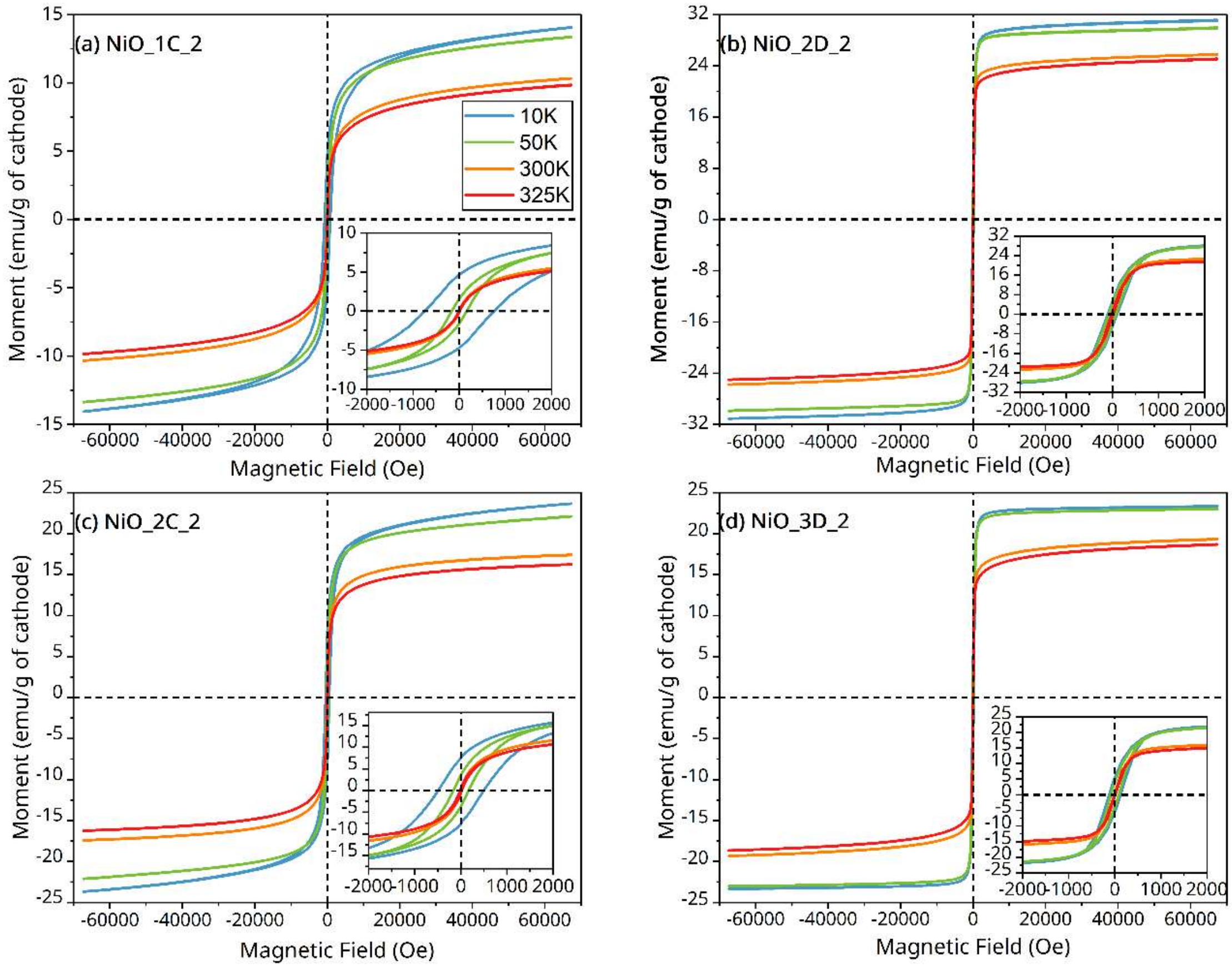

**Fig. S8. Temperature-dependent magnetic hysteresis across independently prepared NiO electrodes.**

M-H loops measured at 10, 50, 300 and 325 K for independently prepared electrodes corresponding to **(a)** NiO_1C_2, **(b)** NiO_2D_2, **(c)** NiO_2C_2 and **(d)** NiO_3D_2.

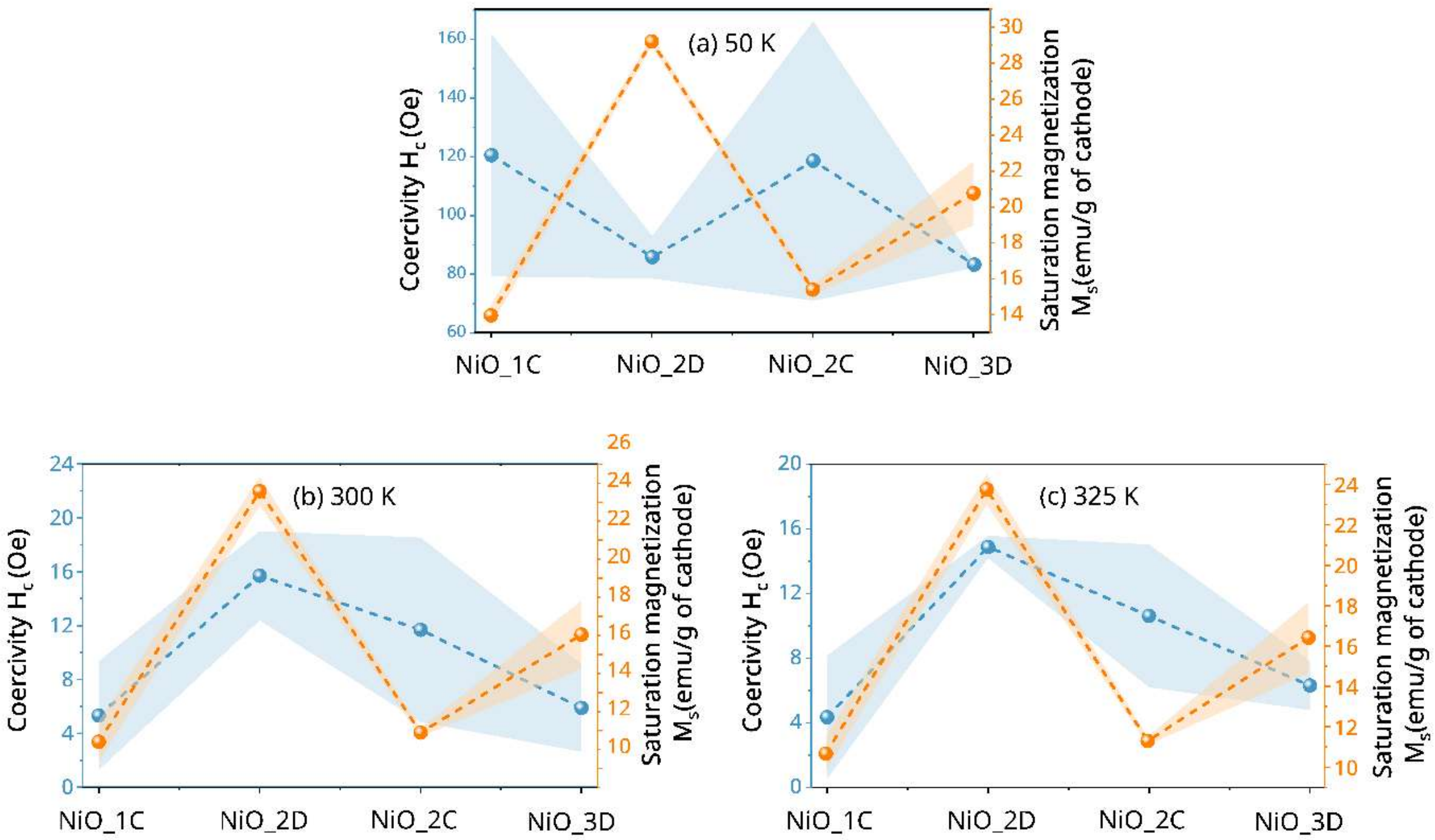

**Fig. S9. Evolution of magnetic properties with electrochemical state.**

Summary of the coercivity and saturation magnetization extracted from two independently prepared samples (**Fig. S7,8**) of NiO_1C, NiO_2D, NiO_2C and NiO_3D measured at **(a)** 50 K, **(b)** 300 K and **(c)** 325 K.

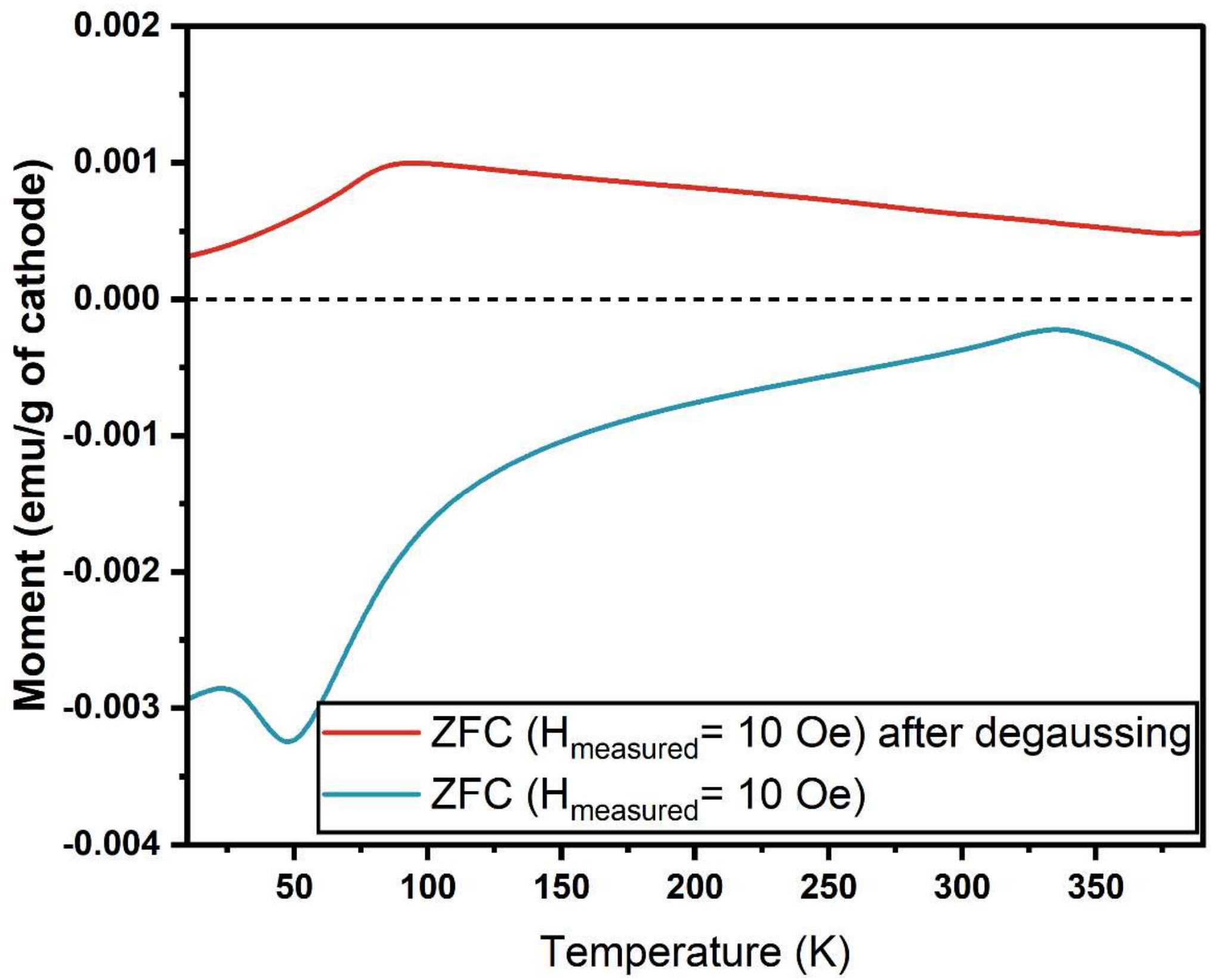

**Fig. S10. Effect of residual trapped field on low-field ZFC magnetization measurements.**
ZFC magnetization measured at 10 Oe before and after degaussing the superconducting magnet (in SQUID). Without degaussing, the measurement exhibits an apparent negative magnetization associated with residual trapped field in the superconducting magnet. Following degaussing, the expected positive magnetization response is observed, demonstrating the sensitivity of low-field ZFC measurements to the residual field of the magnet.

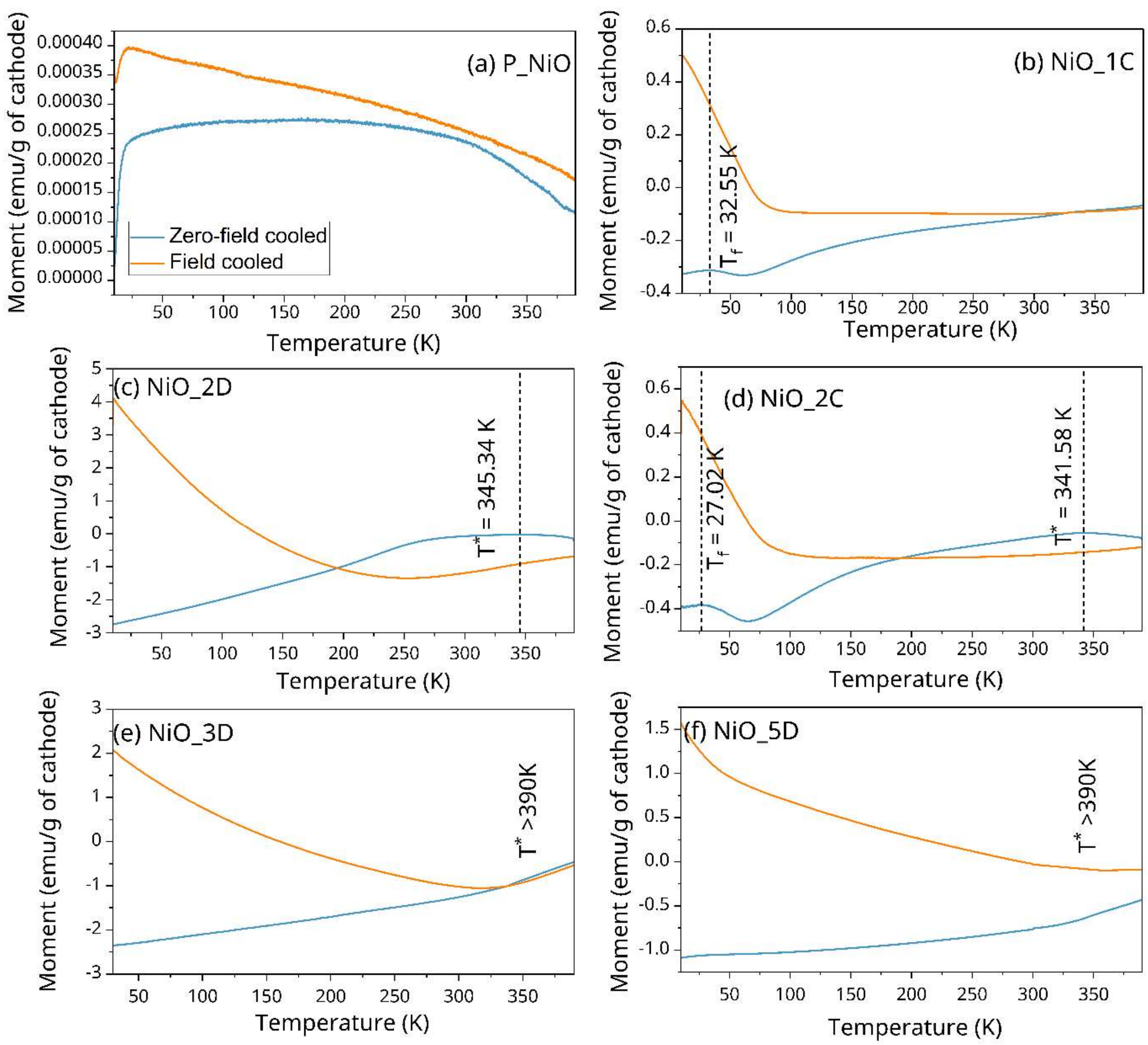

**Fig. S11. Low-field anomalies produced by residual trapped field in the superconducting magnet.**

Temperature-dependent magnetization measurements acquired at 10 Oe without degaussing the superconducting magnet for **(a)** P_NiO, **(b)** NiO_1C, **(c)** NiO_2D, **(d)** NiO_2C, **(e)** NiO_3D and **(f)** NiO_5D shown in the dataset. The non-degaussed measurements exhibit negative magnetization and apparent temperature-dependent anomalies that are not reproduced following magnet degaussing (**Fig. S15**), identifying these features as artifacts associated with residual trapped field rather than intrinsic magnetic transitions.

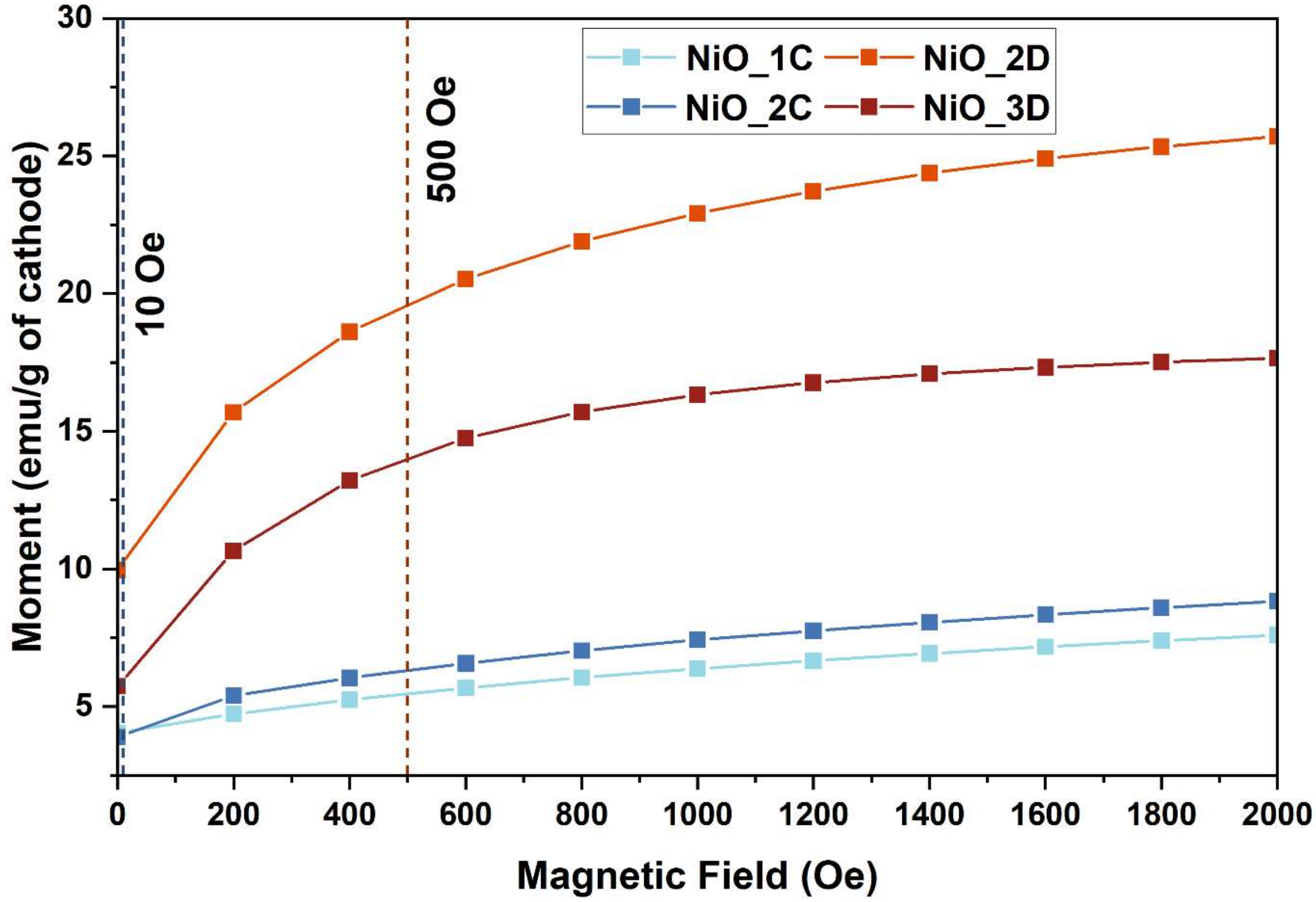

**Fig. S12. Applied fields used for temperature-dependent DC magnetization measurements.** Enlarged first quadrant of the M-H curves of NiO_1C, NiO_2D, NiO_2C and NiO_3D in the low-field region from 0 to 2000 Oe. Dashed vertical lines indicate the 10 and 500 Oe measuring fields used for the temperature-dependent ZFC and FC magnetization measurements.

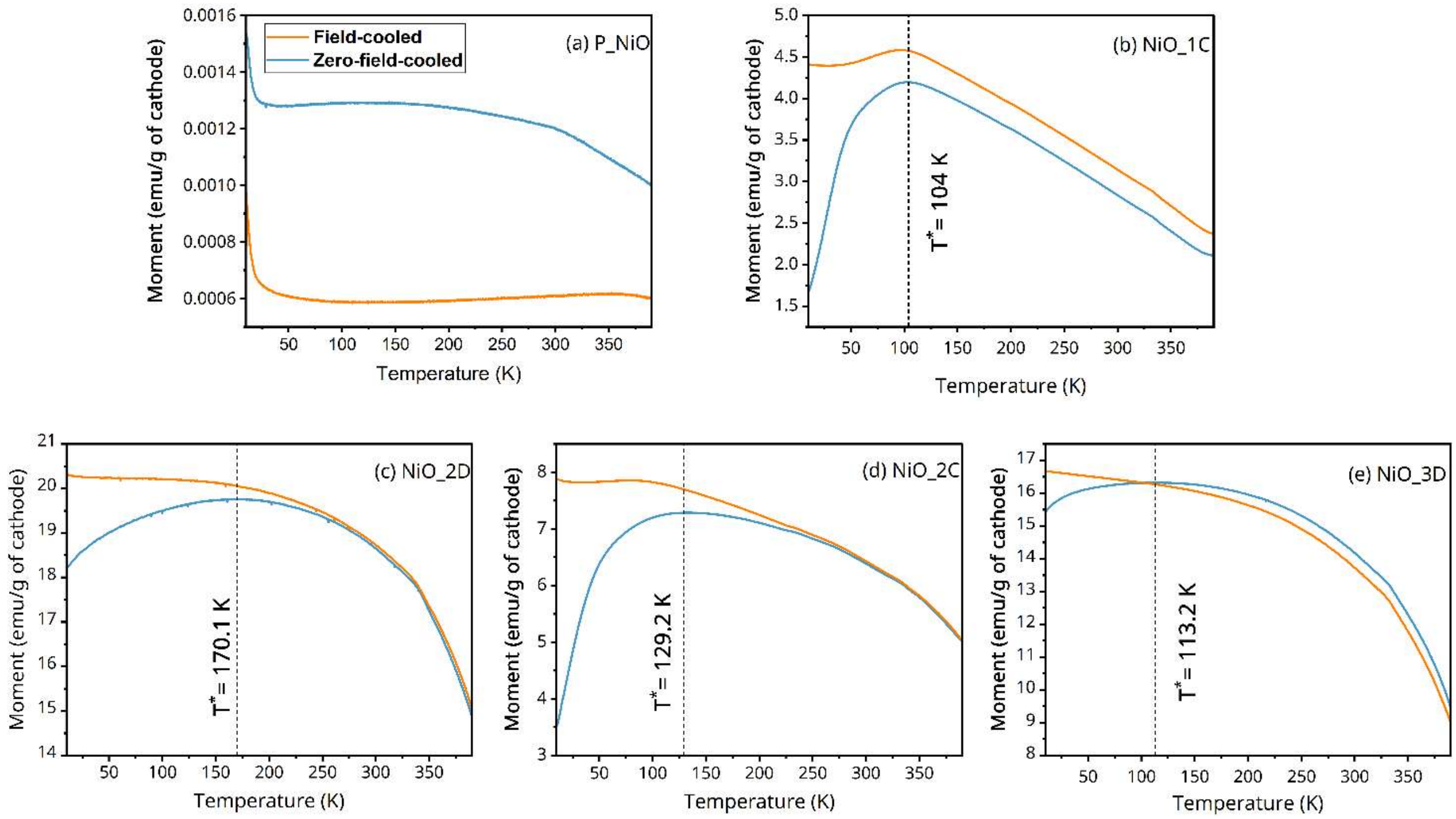

**Fig. S13. Temperature-dependent ZFC and FC magnetization measured at 500 Oe.**
ZFC and FC magnetization measured between 10 and 390 K under an applied field of 500 Oe for **(a)** P_NiO, **(b)** NiO_1C, **(c)** NiO_2D, **(d)** NiO_2C and **(e)** NiO_3D. Dashed vertical lines indicate the characteristic temperatures, $T^*$ identified from the corresponding ZFC curves.

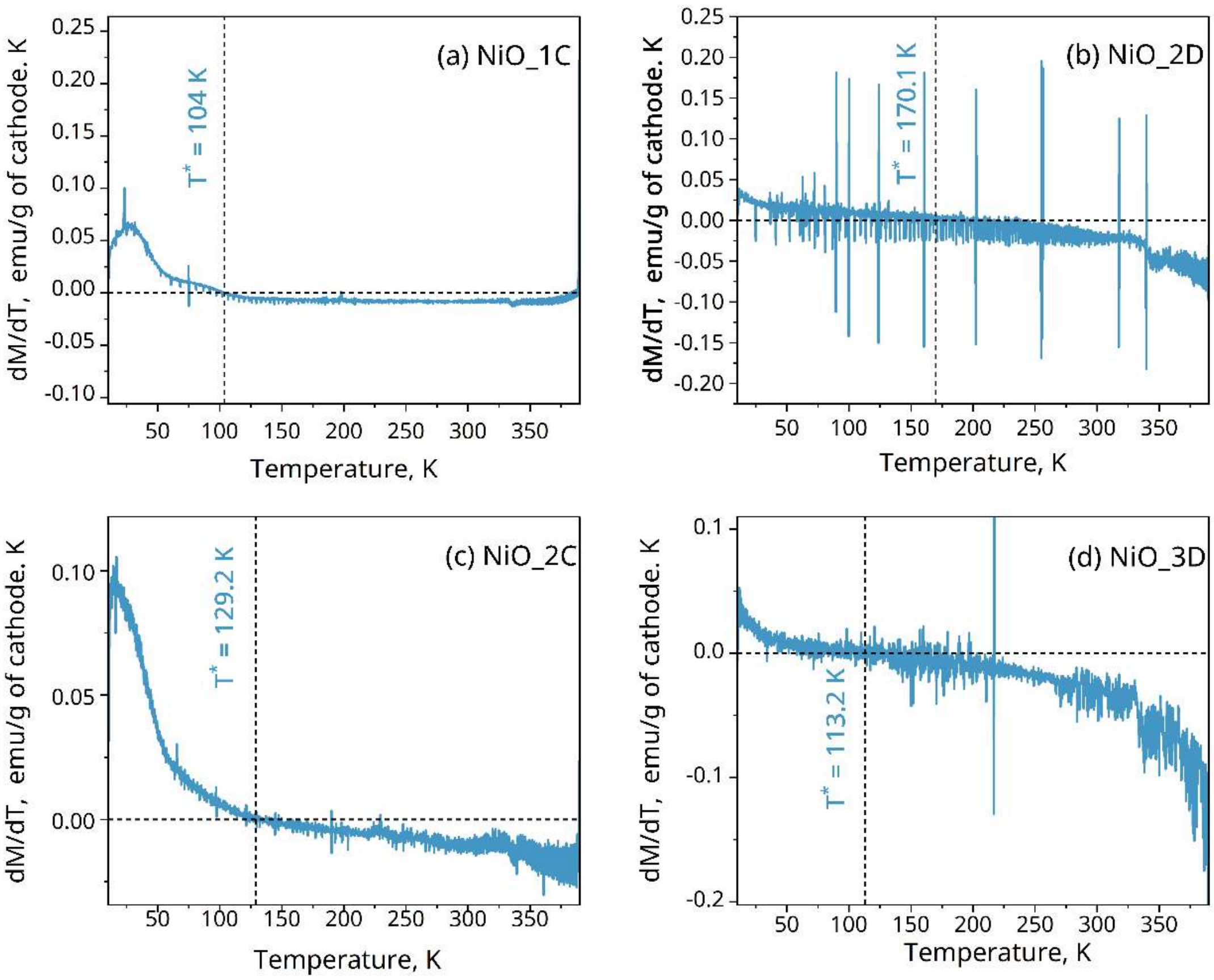

**Fig. S14. Temperature derivatives of the ZFC magnetization measured at 500 Oe.**

First derivative of the ZFC magnetization with respect to temperature, dM/dT, for **(a)** NiO_1C, **(b)** NiO_2D, **(c)** NiO_2C and **(d)** NiO_3D measured under an applied field of 500 Oe. Vertical dashed lines indicate the characteristic temperatures, $T^*$ obtained from the derivative analysis.

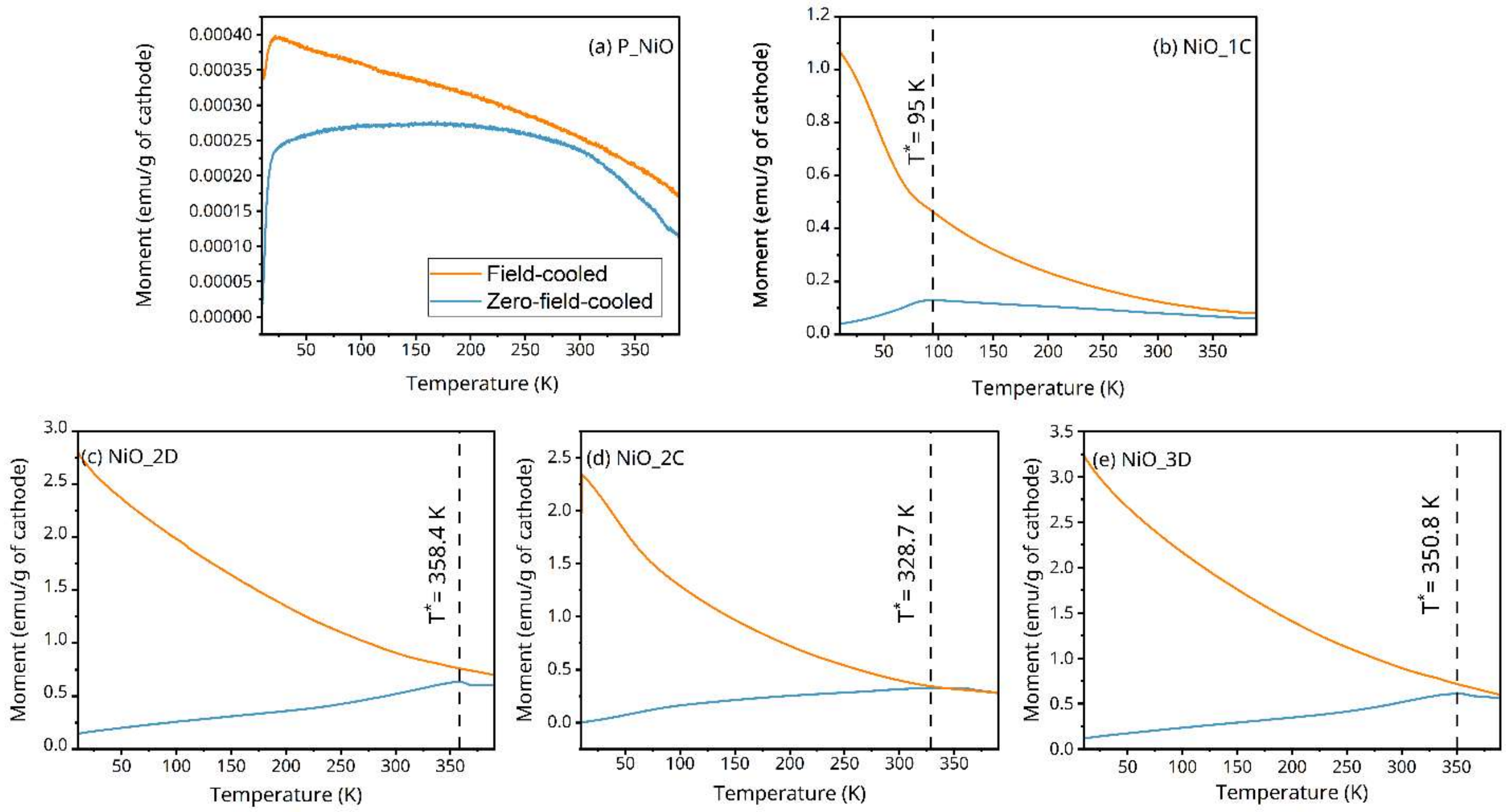

**Fig. S15. Temperature-dependent ZFC and FC magnetization measured at 10 Oe.**

ZFC and FC magnetization measured between 10 and 390 K under an applied field of 10 Oe during measurement (with 100 Oe applied field during cooling) for **(a)** P_NiO, **(b)** NiO_1C, **(c)** NiO_2D, **(d)** NiO_2C and **(e)** NiO_3D. Dashed vertical lines indicate the characteristic temperatures, $T^*$ identified from the corresponding ZFC curves.

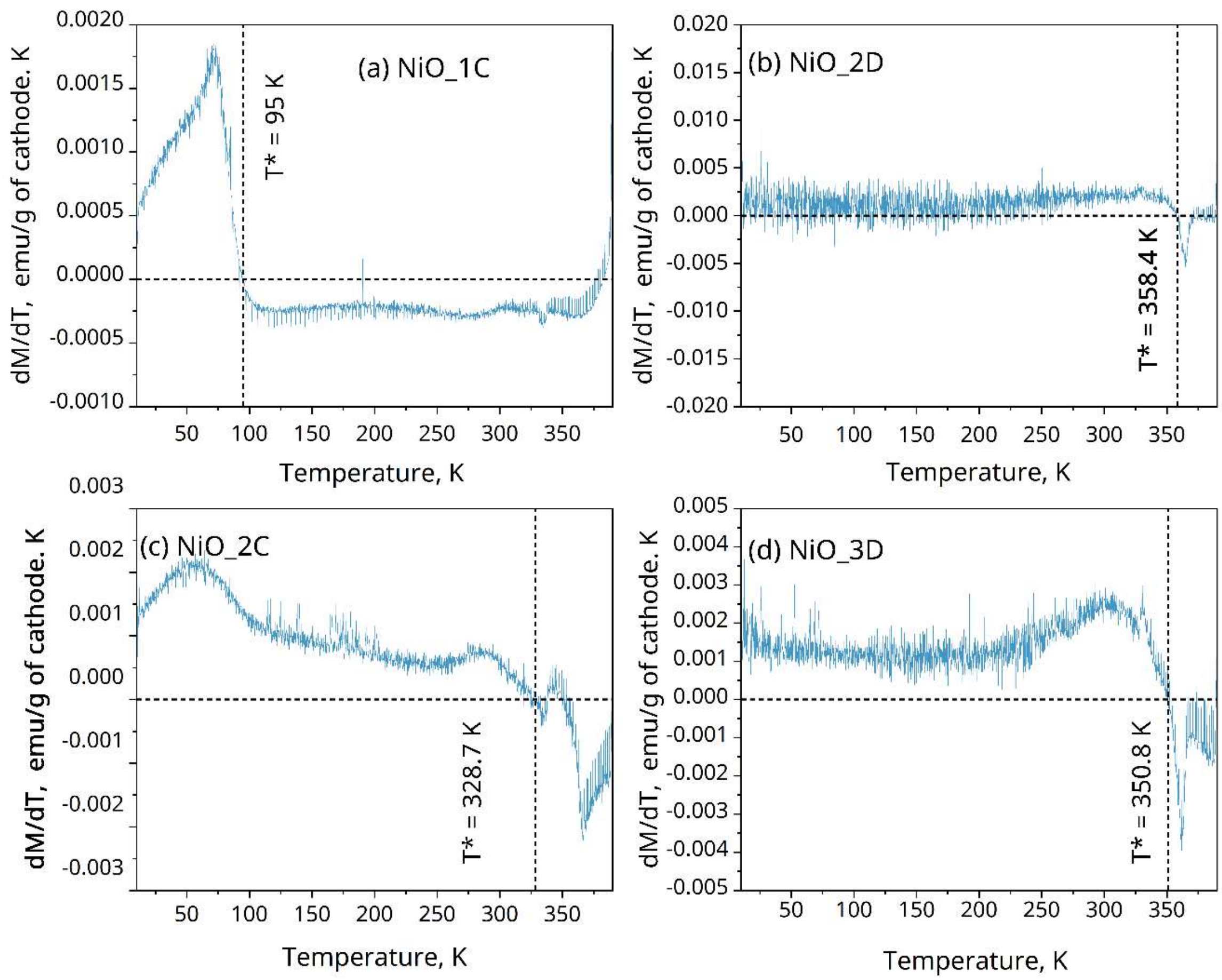

**Fig. S16. Temperature derivatives of the ZFC magnetization measured at 10 Oe.**

First derivative of the ZFC magnetization with respect to temperature, dM/dT, for **(a)** NiO_1C, **(b)** NiO_2D, **(c)** NiO_2C and **(d)** NiO_3D measured under an applied field of 10 Oe. Vertical dashed lines indicate the characteristic temperatures, $T^*$ obtained from the derivative analysis.

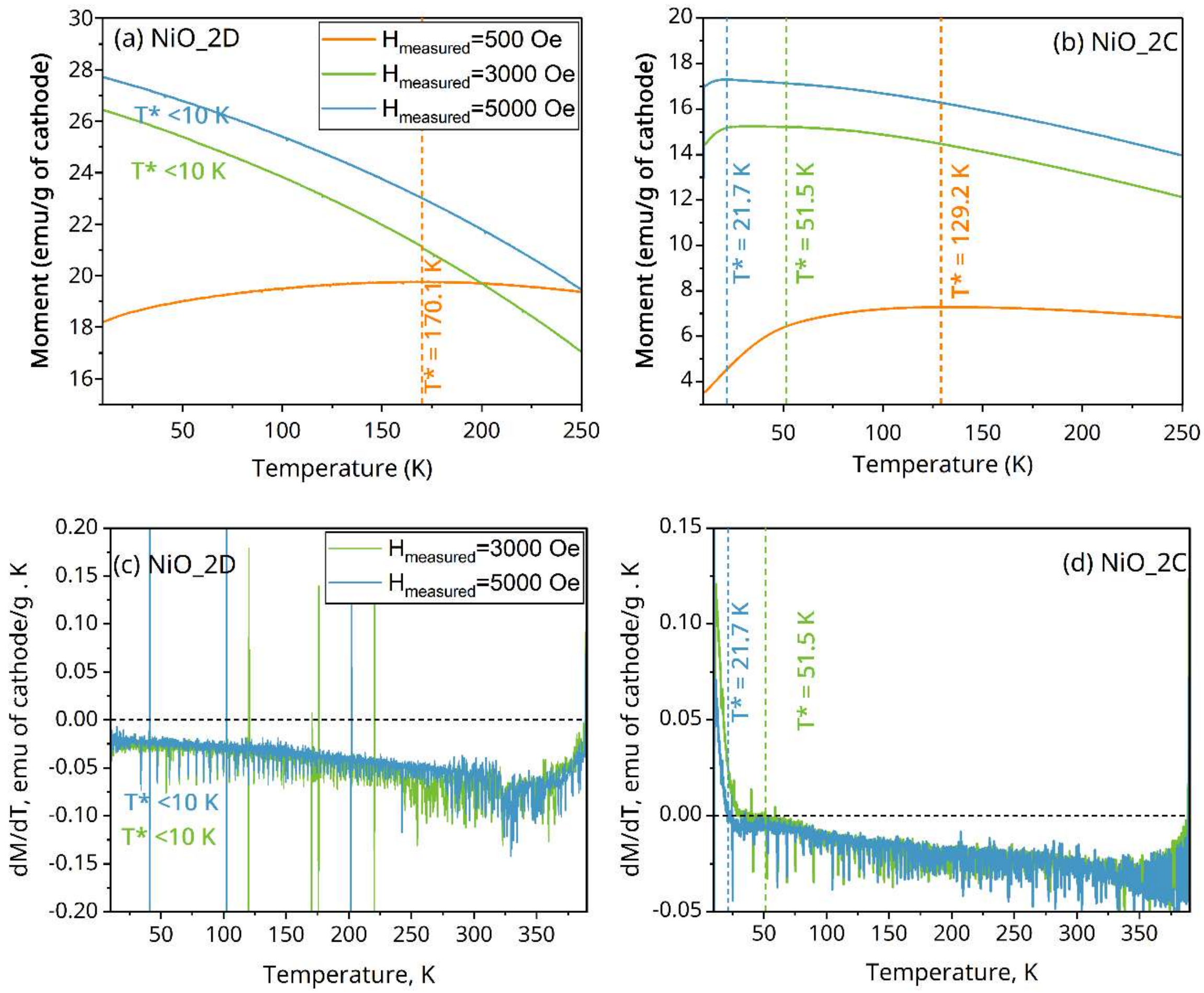

**Fig. S17. Field dependence of the ZFC magnetization of discharged and charged NiO electrodes.**

ZFC magnetization of **(a)** NiO_2D and **(b)** NiO_2C measured under applied fields of 500, 3000 and 5000 Oe. Dashed vertical lines indicate characteristic temperatures, $T^*$ extracted from the ZFC response. Corresponding temperature derivatives, dM/dT, for **(c)** NiO_2D and **(d)** NiO_2C.

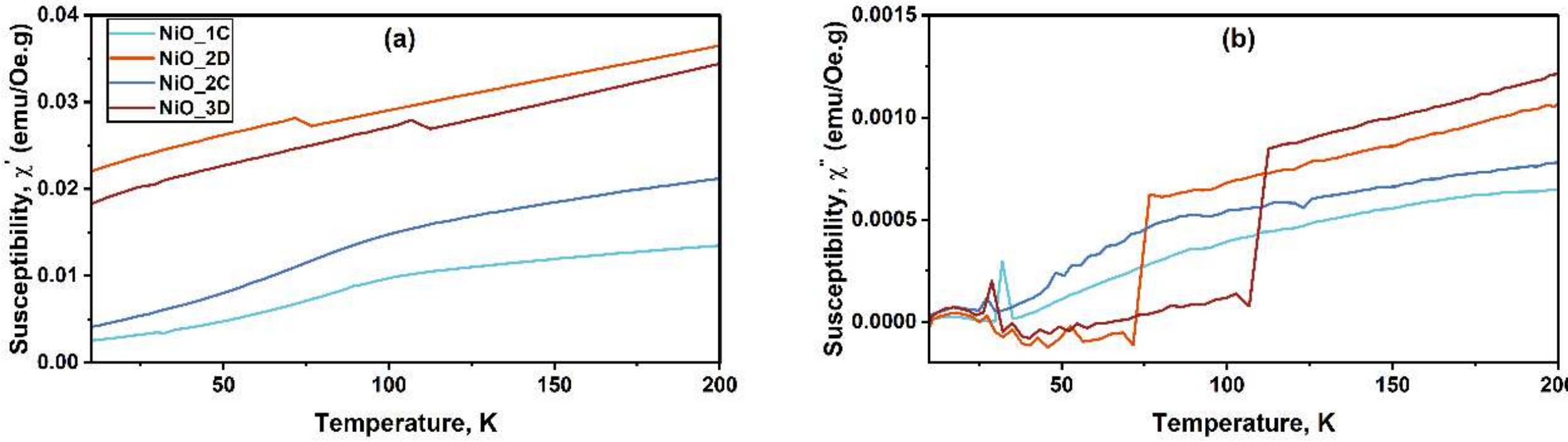

**Fig. S18. AC magnetic susceptibility under a 10 Oe DC bias field.**

Temperature-dependent AC magnetic susceptibility of NiO_1C, NiO_2D, NiO_2C and NiO_3D measured using an AC excitation of 2 Oe at 1000 Hz with a DC bias field of 10 Oe. **(a)** Real component of the AC susceptibility, $\chi'$; **(b)** imaginary component, $\chi''$ as a function of temperature. Low-temperature features common to both sample and background measurements, together with a high-frequency $\chi''$ baseline offset observed only in the discharged electrodes, are attributed to instrumental or background contributions and are excluded from physical interpretation.

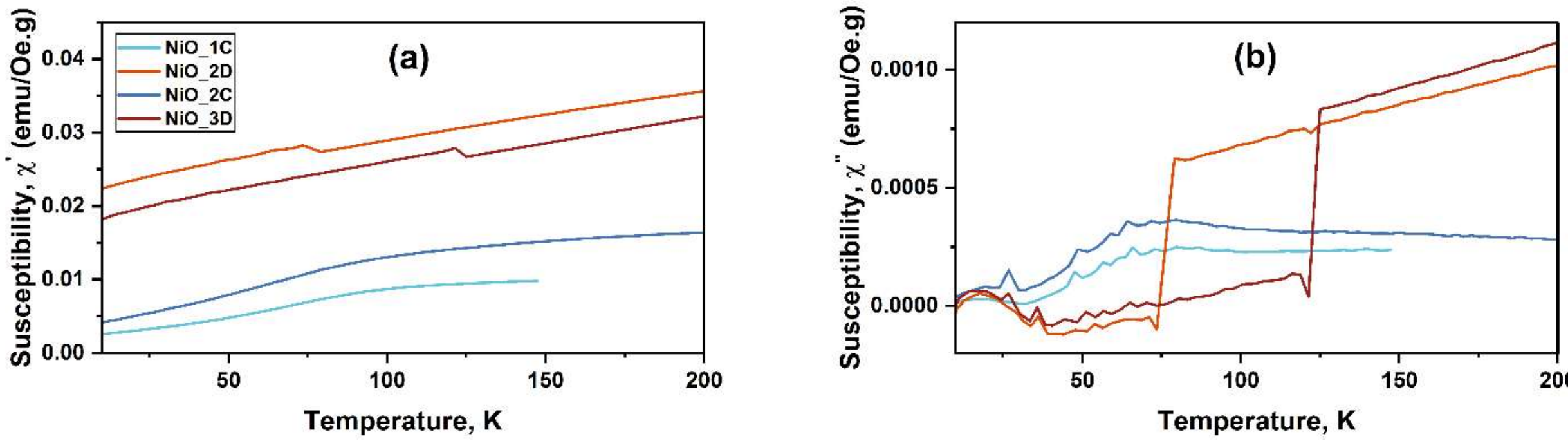

**Fig. S19. AC magnetic susceptibility under a 100 Oe DC bias field.**

Temperature-dependent AC magnetic susceptibility of NiO_1C, NiO_2D, NiO_2C and NiO_3D measured using an AC excitation of 2 Oe at 1000 Hz with a DC bias field of 100 Oe. **(a)** Real component of the AC susceptibility, $\chi'$; **(b)** imaginary component, $\chi''$ as a function of temperature. Low-temperature features common to both sample and background measurements, together with a high-frequency $\chi''$ baseline offset observed only in the discharged electrodes, are attributed to instrumental or background contributions and are excluded from physical interpretation.

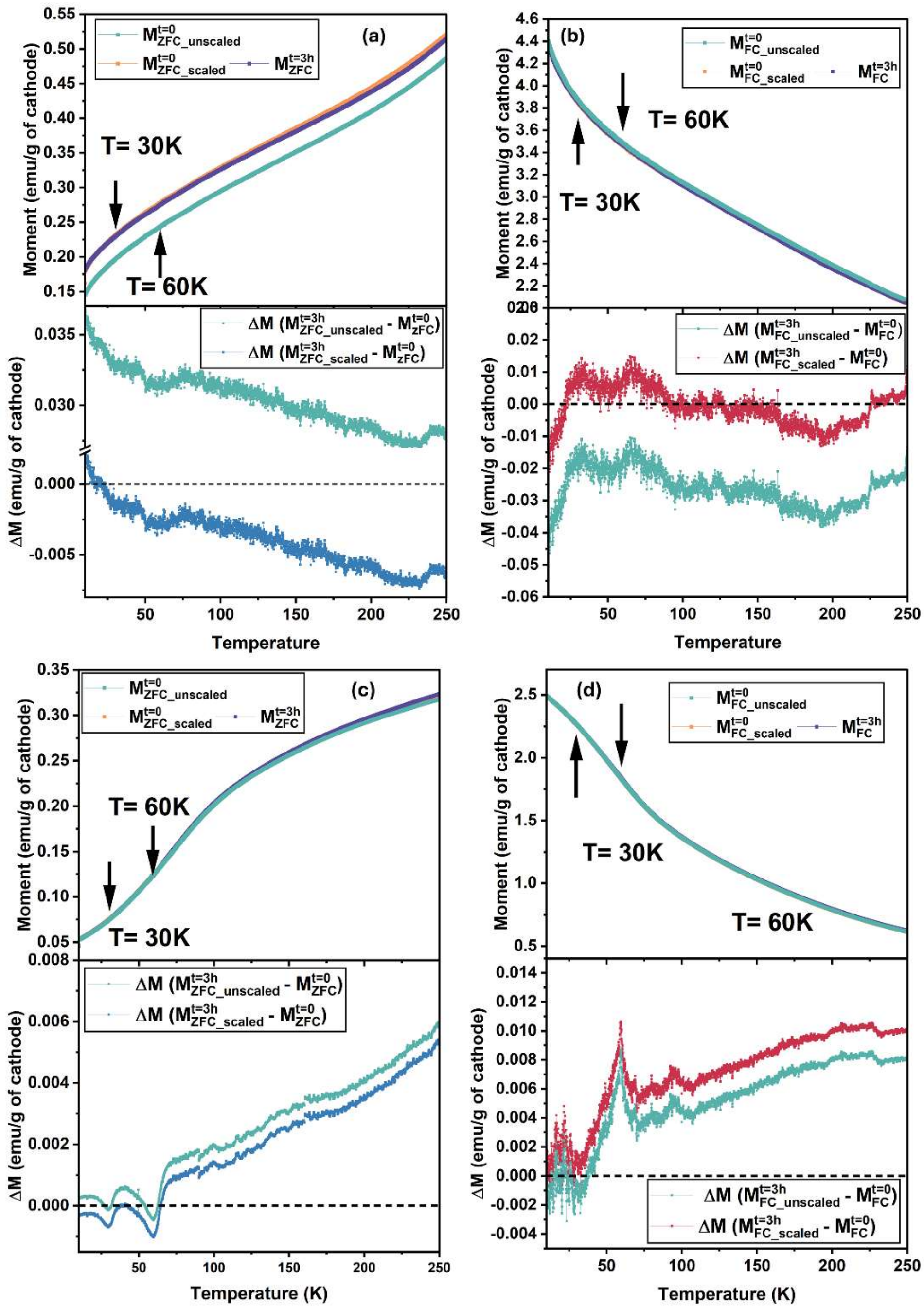

**Fig. S20. ZFC and FC magnetic memory of discharged and charged NiO electrodes.**

ZFC and FC memory measurements performed with intermediate temperature stops at 30 and 60 K. **(a,b)** Memory measurements for NiO_2D and **(c,d)** NiO_2C. Upper panels compare the reference magnetization curves with those obtained following the stop-and-wait protocol, with arrows marking the stop temperatures. Lower panels show the corresponding magnetization difference, ΔM, between the memory and reference curves.

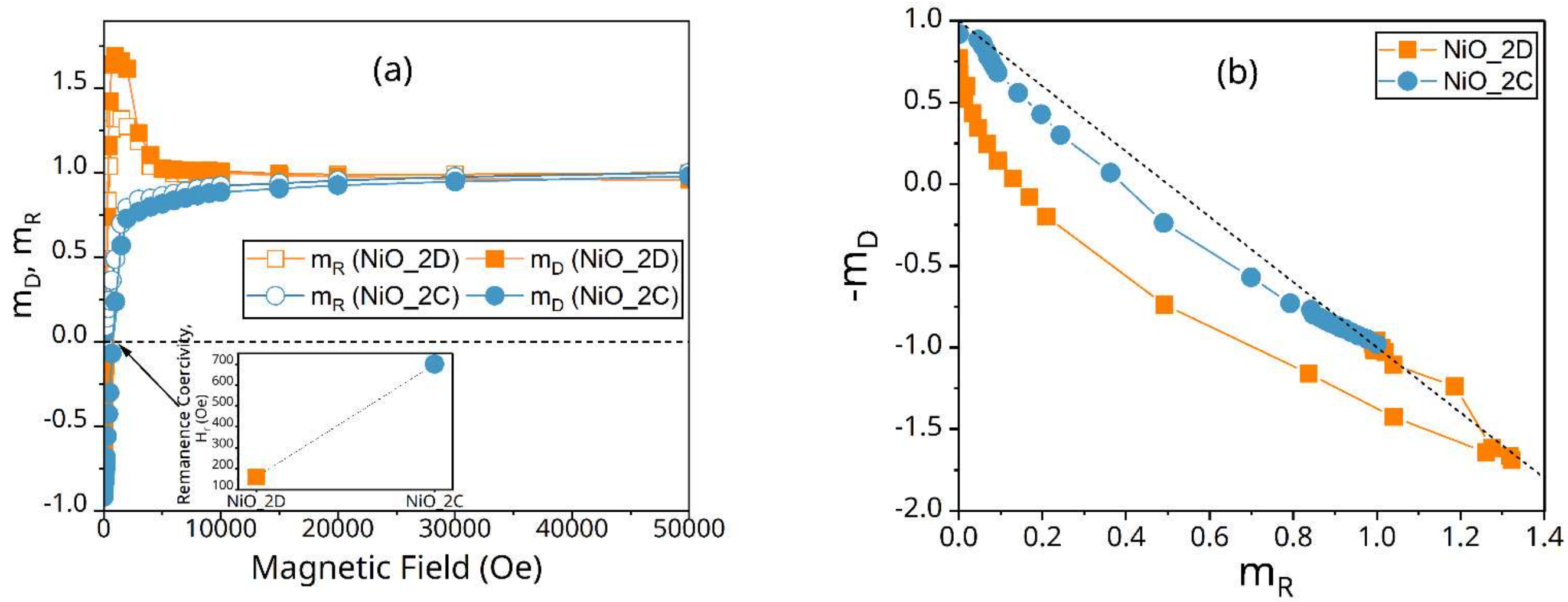

**Fig. S21. Henkel analysis reveals distinct interaction pathways in electrochemically programmed magnetic states.**

**(a)** Normalized isothermal remanent magnetization ($m_R$) and dc demagnetization remanence ($m_D$) measured at 10 K for NiO_2D and NiO_2C electrodes as a function of applied magnetic field. Inset: Extracted remanence coercivity, $H_r$.

**(d)** Extended Henkel plot ($-m_D$ versus $m_R$) where the dashed line denotes the Wohlfarth relation expected for non-interacting single-domain particles.

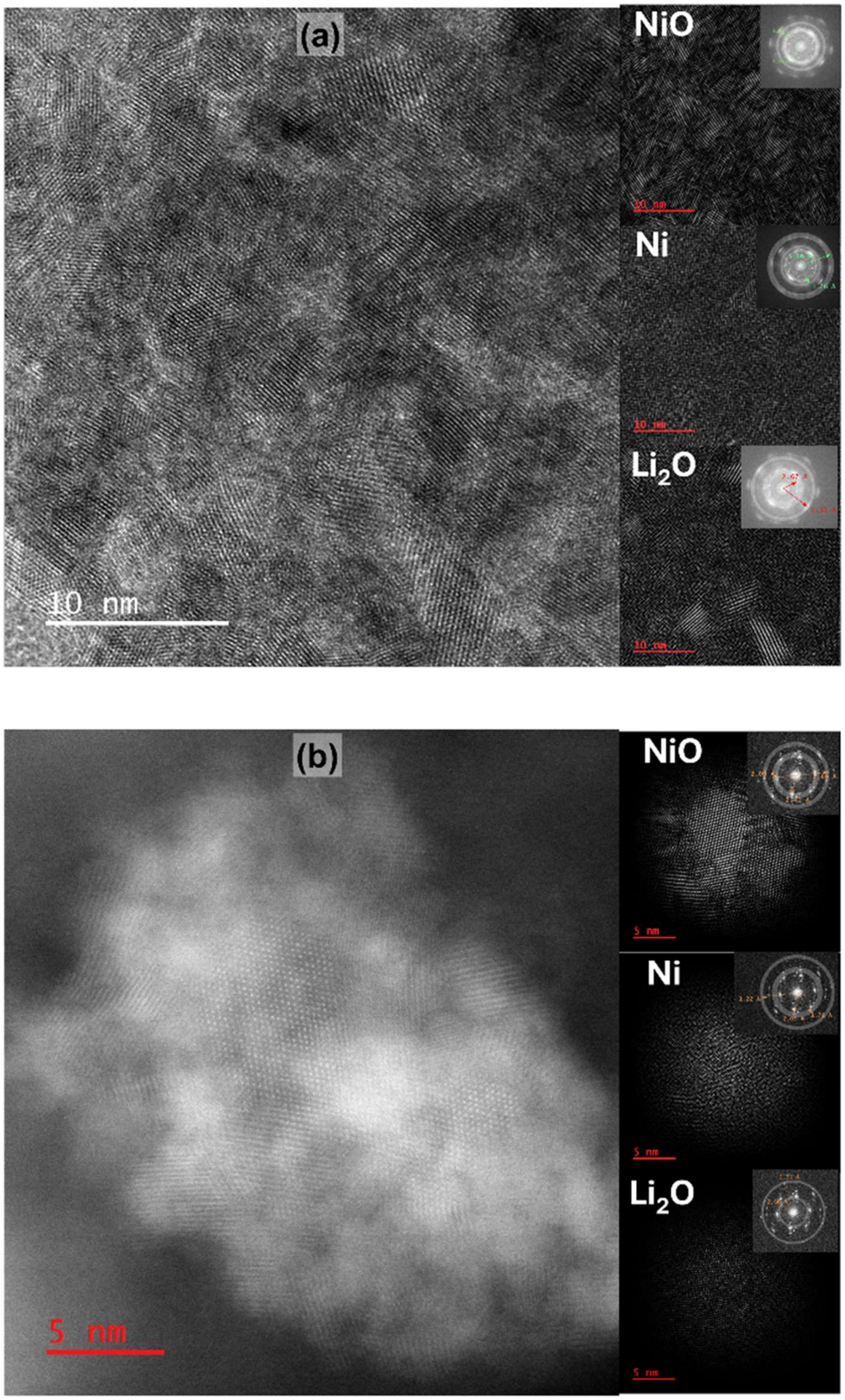

**Fig. S22. Evolution of magnetic properties with electrochemical state.**
DF-STEM of **(a)** NiO_2D and **(b)** NiO_2C with fourier-filtered images corresponding to reflections assigned to NiO, Ni and $Li_2O$.

| Sample | Ni crystallite diameter (nm) | Blocking temperature, $T_B$(K) |
|---|---|---|
| **NiO_1C** | 2.32 | 0.1 |
| **NiO_2D** | 2.37 | 0.1 |
| **NiO_2C** | 2.06 | 0.07 |
| **NiO_3D** | 5.00 | 0.95 |

**Table S1. Blocking temperature assuming isolated Ni crystallites for charged and discharged samples**

| Sample | Saturation magnetization, $M_s$ | Remanence magnetization, $M_r$ | Squareness, $M_r/M_s$ |
|---|---|---|---|
| **NiO_1C** | 14.1 | 4.4 | 0.31 |
| **NiO_2D** | 29.9 | 8 | 0.27 |
| **NiO_2C** | 16 | 5.9 | 0.37 |
| **NiO_3D** | 21.1 | 6 | 0.28 |

**Table S2. Extracted average saturation and remanence magnetization at 10K for charged and discharged samples**